\documentclass[twocolumn,showpacs,preprintnumbers,amsmath,amssymb]{revtex4-1}

\pdfoutput=1
\usepackage[pdftex]{graphicx}
\usepackage[caption=false]{subfig}% For figures side by side
\usepackage[update,prepend]{epstopdf}

\usepackage{dcolumn}% Align table columns on decimal point\right]
\usepackage{bm}% bold math

\usepackage{color}
\usepackage{amstext}

\begin{document}

\title{Spin and charge density waves in the Lieb lattice}

\author{J. D. Gouveia, R. G. Dias}
\affiliation{Departamento de F\'{\i}sica, I3N, Universidade de Aveiro, Campus de Santiago, Portugal}%

\date{\today}

\begin{abstract}
We study the mean-field phase diagram of the two-dimensional (2D) Hubbard model in the Lieb lattice allowing for spin and charge density waves. Previous studies of this diagram have shown that the mean-field magnetization surprisingly deviates from the value predicted by Lieb's theorem \cite{Lieb1989} as the on-site repulsive Coulomb interaction ($U$) becomes smaller \cite{Gouveia2015}.
Here, we show that in order for Lieb's theorem to be satisfied, a more complex mean-field approach should be followed in the case of bipartite lattices or other lattices whose unit cells contain more than two types of atoms.
In the case of the Lieb lattice, we show that, by allowing the system to modulate the magnetization and charge density between sublattices, the difference in the absolute values of the magnetization of the sublattices, $m_{\text{Lieb}}$, at half-filling, saturates at the exact value $1/2$ for any value of $U$, as predicted by Lieb. Additionally, Lieb's relation, $m_{\text{Lieb}}=1/2$, is verified approximately for large $U$, in the $n \in [2/3,4/3]$ range. This range includes not only the ferromagnetic region of the phase diagram of the Lieb lattice (see Ref.~\onlinecite{Gouveia2015}), but also the adjacent spiral regions. In fact, in this lattice, below or at half-filling, $m_{\text{Lieb}}$ is simply the filling of the quasi-flat bands in the mean-field energy dispersion both for large and small $U$.
\end{abstract}


\maketitle

\section{Introduction}
\label{}

Despite intense research in the last few decades, the 2D Hubbard model in the square lattice has remained an open theoretical problem in the field of the strong correlated systems \cite{Bednorz1986,RevModPhys.70.897}. Although it is known that at half-filling, the spin dynamics of the 2D Hubbard model is described by the Heisenberg antiferromagnetic exchange term \cite{Dias1992}, there is no consensus regarding the ground state  magnetic phase diagram of this model. In fact, even at the mean-field (MF) level, depending on the magnetic phases allowed, different authors obtain different diagrams for the square lattice \cite{Marder2000}. The  traditional orderings are ferromagnetism, antiferromagnetism and paramagnetism \cite{PhysRev.142.350,J.Dorantes-Davila1083,E.Kaxiras1988,Coppersmith1989,A.Richter1978}. Later, spin spiral phases, a generalization of the previous three, were introduced \cite{S.Sarker1991}. The MF phase diagram became even more complex with the consideration of spatial phase separation \cite{Langmann2007,P.A.Igoshev2010,Schumacher1983}.

\begin{figure}
\begin{minipage}{.5 \columnwidth}
    \subfloat[]{\label{fig-Fsublattice_AFglobal}\includegraphics[width=.8 \textwidth]{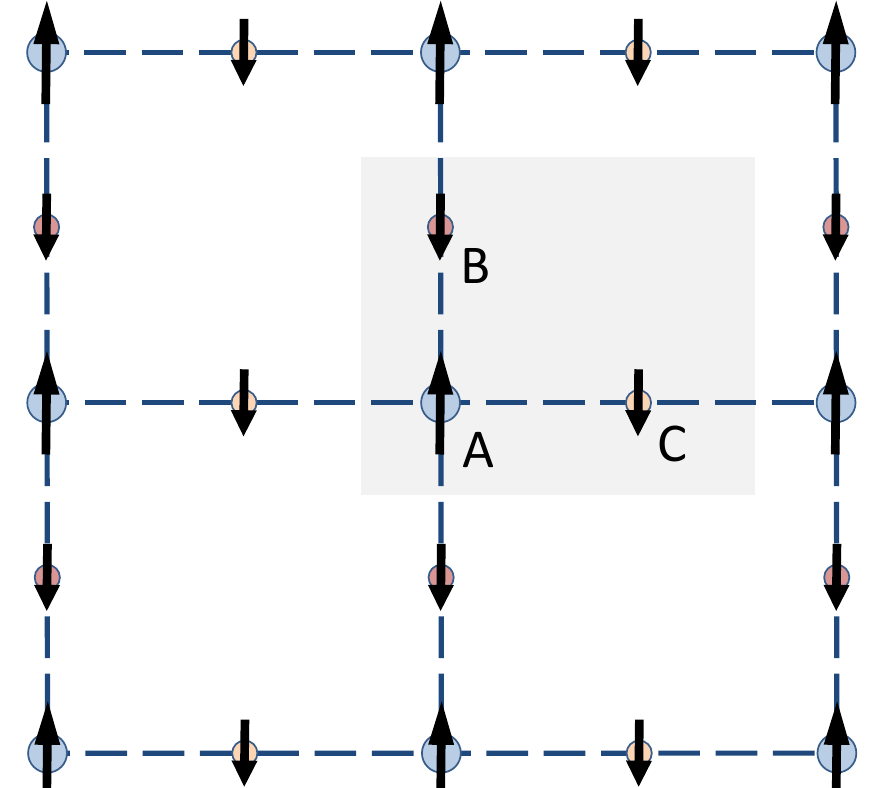}}  \\
    \subfloat[]{\label{fig-Lieb_tb_E_and_n}\includegraphics[width=1.\textwidth]{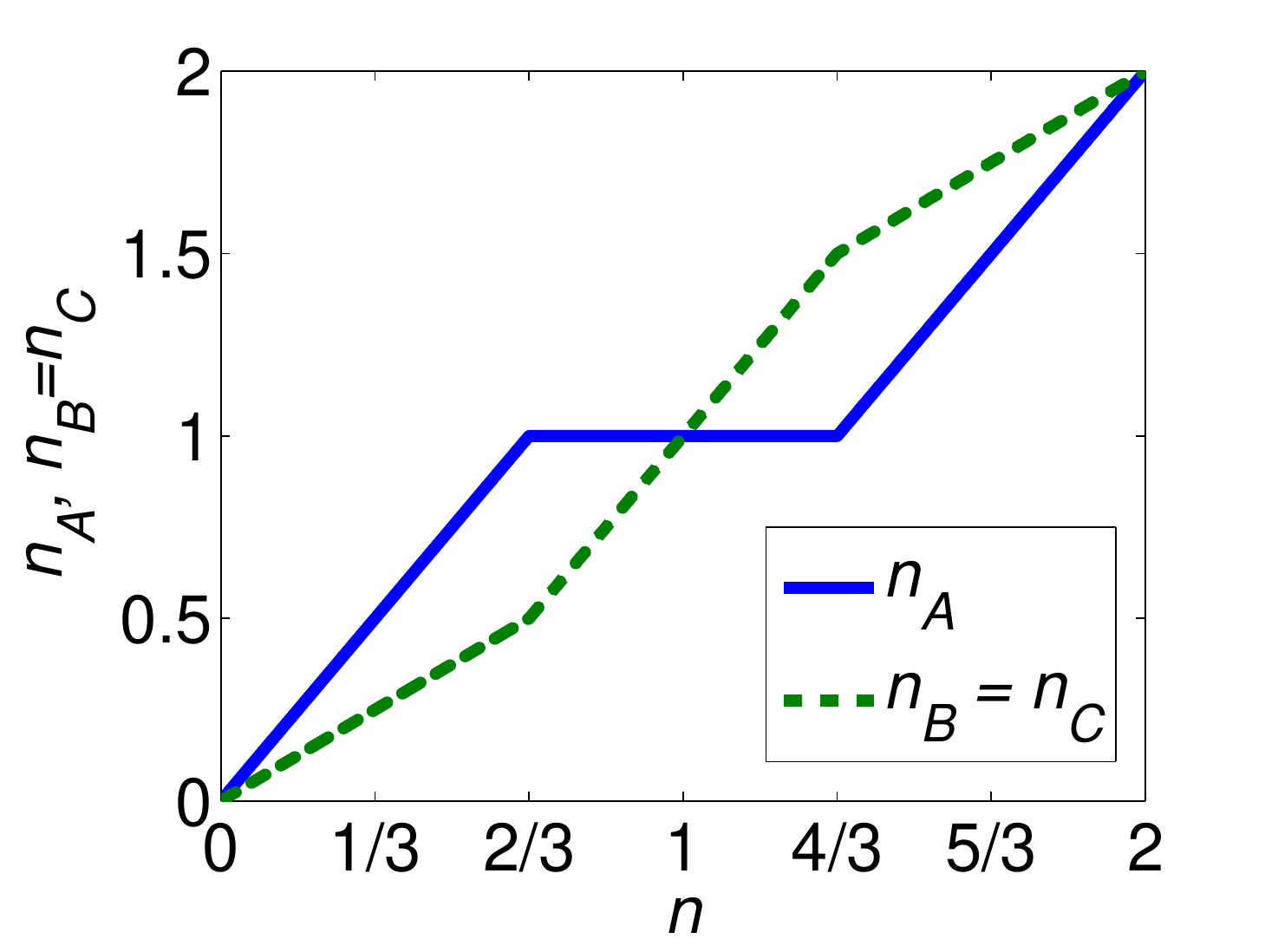}}
\end{minipage}
\hspace{-.4cm}
\begin{minipage}{.4 \columnwidth}
    \subfloat[]{\label{fig-Lieb_tb_dispersionrelation}\includegraphics[height=5cm]{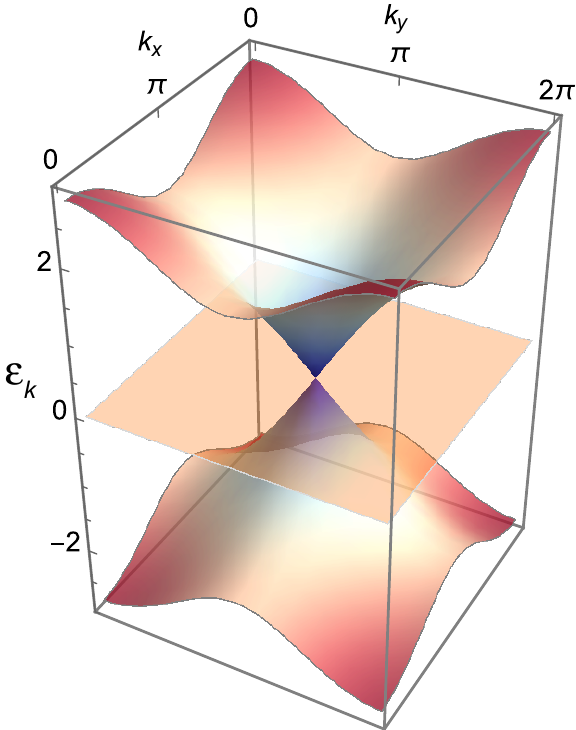}}  \\
\end{minipage}
\caption{(a) The Lieb lattice is a line-centered square lattice, comprising three sublattices, A, B, and C and having one atom of each type in a unit cell. The circles represent atomic nuclei and the arrows represent spins. At half-filling, the Lieb lattice is ferromagnetic within each sublattice but antiferromagnetic overall. Moreover, at half-filling, the total spin per unit cell is $1/2$, as predicted by Lieb \cite{Lieb1989}. (b) Plot of the tight-binding ($U=0$) particle density of each sublattice of the Lieb lattice, A, B or C, as a function of the total particle density. (c) Plot of the tight-binding dispersion relation $\varepsilon (k_x,k_y)$ of the Lieb lattice. The flat band is made up entirely of B and C orbitals.}
\label{fig-Lieb_intro}
\end{figure}

The Hubbard model in decorated 2D lattices has also been extensively studied, motivated by the search for metallic (flat-band) ferromagnetism. These decorated lattices fall into three categories: the Lieb \cite{Lieb1989}, Mielke \cite{Mielke1992} and Tasaki lattices \cite{Tasaki1992}. All of these lattices share a common feature: the presence of flat bands in the energy dispersion relation. In the particular case of the Lieb's lattices, the flat bands result from the topology of the lattice, while in the case of Mielke and Tasaki lattices, the flat bands reflect longer-range transfer integrals in the system. One of the most representative examples of decorated lattices is the Lieb lattice, which can be obtained from the 2D square lattice, for example, by inserting an extra atom between every two nearest-neighbours (see Fig.~\ref{fig-Fsublattice_AFglobal}). Each unit cell (shaded rectangle in Fig.~\ref{fig-Fsublattice_AFglobal}) has 3 atoms, one of each kind: A, B and C, whose relative occupation is depicted in Fig.~\ref{fig-Lieb_tb_E_and_n}, in the limit of no interactions. Fig.~\ref{fig-Lieb_tb_dispersionrelation} shows the energy bands of the Lieb lattice in this limit. Examples of materials whose structure resembles the Lieb lattice include La$_{2-x}$Sr$_x$CuO$_4$ and YBa$_2$Cu$_3$O$_7$, two well-known high-$T_c$ superconductors with weakly coupled CuO$_2$ planes \cite{Emery1987,Scalettar1991}.

A theorem by Lieb \cite{Lieb1989} states that, in the particular case of bipartite lattices (i.e., lattices with two sublattices, A and B, such that each site on sublattice A has its nearest neighbors on sublattice B, and vice versa), the ground state is ferromagnetic at half-filling ($n=1$, or one electron per lattice site), as long as the number of atoms of each sublattice is different. However, for example in the case of the Lieb lattice (a line-centered square lattice \cite{Wang2014}), this ground state should be identified with ferrimagnetism \cite{Mielke1993}. In fact, although each sublattice is indeed ferromagnetic, there is antiferromagnetic ordering between every pair of nearest neighbours \cite{Gouveia2015} (see Fig.~\ref{fig-Fsublattice_AFglobal}).

\begin{figure}[tb]
\centering
\includegraphics[height=5.2cm]{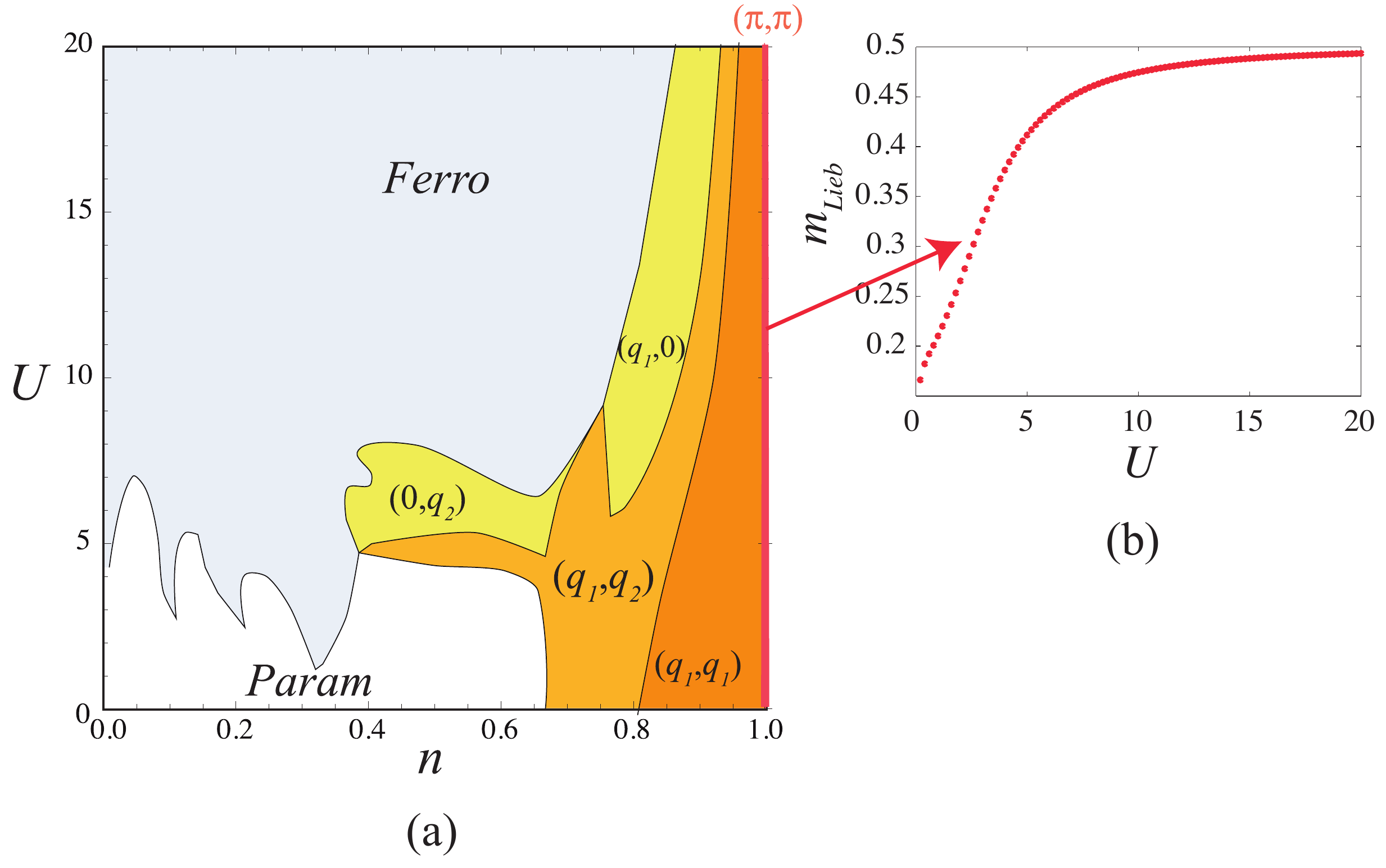}
\caption{(a) Mean-field magnetic phase diagram of the Hubbard model in the Lieb lattice, obtained in Ref.~\onlinecite{Gouveia2015} and (b) difference in the absolute values of the magnetization of the sublattices,  $m_{\text{Lieb}} = (|m_B|+|m_C|-|m_A|)/2$, of the Lieb lattice at half-filling, using the mean-field results from the same reference, where the on-site magnetization is assumed to be the same on all sublattices.}
\label{fig-diagram_standard_Lieb}
\end{figure}

The magnetic phase diagram of the Hubbard model in the Lieb lattice was recently studied by us \cite{Gouveia2015}. We showed that the mean-field magnetization per unit cell at half-filling, $m_{\text{Lieb}}=(|m_B|+|m_C|-|m_A|)/2$, assuming  the particle density ($n$)  of the tight-binding limit (see Fig.~\ref{fig-Lieb_tb_E_and_n}) and the same magnetization ($m$) in the whole lattice, surprisingly deviates from the value predicted by Lieb's theorem \cite{Lieb1989} as the on-site repulsive Coulomb interaction ($U$) becomes smaller, although these two assumptions are common in mean-field studies  \cite{Noda2014,Gouveia2014,Dzierzawa1992,Langmann1997,Gouveia2015}.  Lieb's theorem predicts that the magnetization per unit cell, $m_{\text{Lieb}}$, is $1/2$ for any $U$ at half-filling. Fig.~\ref{fig-diagram_standard_Lieb} shows both the mean-field phase diagram of the Hubbard model in the Lieb lattice, and the value of $m_{\text{Lieb}}$, using the mean-field results from Ref.~\onlinecite{Gouveia2015}. For the results to agree with Lieb's theorem, one should have $m_{\text{Lieb}} = 1/2$ for any $U$. Although in the strong coupling limit ($U \gg t$), the mean-field result satisfies Lieb's theorem, it is far from correct near the tight-binding limit ($U=0$).

In this manuscript, we study the magnetic phase diagram of the  Lieb lattice allowing for different average occupations ($n_A$, $n_B$, and $n_C$) and magnetization amplitudes ($m_A$, $m_B$, and $m_C$) in each sublattice. We find that with these new considerations, Lieb's relation, $m_{\text{Lieb}}=1/2$,  is satisfied for any $U$ at half-filling, and satisfied approximately for large $U$, in the $n \in [2/3,4/3]$ range, which includes not only the ferromagnetic region of the phase diagram of the Lieb lattice in Fig.~\ref{fig-diagram_standard_Lieb}, but also the adjacent spiral regions (note that, away from half-filling, $m_{\text{Lieb}}$ is no longer the unit cell magnetization, but gives the difference in the absolute values of the magnetization of the sublattices).
An important point in this result is that finite $m_{\text{Lieb}}$  reflects the existence of quasi-flat bands in the mean-field energy dispersion. These quasi-flat bands are present, not only for small $U$, but also for large $U$. In fact, below or at half-filling, $m_{\text{Lieb}}$ is approximately the filling of the flat bands both for large and small $U$.

Our mean-field approach follows that of Bach and Poelchau \cite{Bach1996} (see also Refs. \onlinecite{Langmann2007,Bach1994}). The formalism with further mathematical details can be found in Ref. \onlinecite{Woul2007}.

The organization of this paper is as follows. We begin by presenting some key results for the Hubbard model and mean-field. Secondly, we revisit the tight-binding limit of the Lieb lattice. We then proceed to adding electronic interactions to the Hamiltonian and calculating its mean-field counterpart. Finally, we show our results, discuss their meaningfulness, compare them to analytical calculations and take conclusions. In Appendix A, we briefly outline the derivation which leads to the results in section II, and in Appendix B, we explain in more detail our method of calculating saddle points, which we used to obtain the results in section V.

\section{Mean-field method for the Hubbard model}

In this section, we adapt a key result presented as Theorem 4.14 in Ref.~\onlinecite{Woul2007}. This derivation was first done by Lieb and his colaborators \cite{Bach1994} and simplified by Bach and Poelchau \cite{Bach1996}; in Appendix A we show an adaptation of the derivation in Ref.~\onlinecite{Bach1996}. Alternative approaches can be found in Refs. \onlinecite{Langmann1997} and \onlinecite{Langmann1997a}. The result consists of the following abridged derivation.

We begin with the Hubbard Hamiltonian, given by
\begin{equation}
H = t \sum\limits_{\langle x,y \rangle , \sigma} c_{x,\sigma}^{\dagger} c_{y,\sigma} + U \sum\limits_{x} \hat{n}_{x,\uparrow} \hat{n}_{x,\downarrow} .
\label{eq-H_calculateOmega}
\end{equation}
Here, $t$ is the hopping parameter between nearest neighbours and $c_{x,\sigma}^{\dagger}$ ($c_{y,\sigma}$) is the creation (annihilation) operator of an electron on site $x$ ($y$) with spin $\sigma = \uparrow , \downarrow$. The letters $x$ and $y$ denote lattice sites, $\langle x,y \rangle$ stands for nearest neighbors and $U$ is the on-site repulsive. The total number of particles is $N$.

With the intent of finding the mean-field Helmholtz free energy, $F_{\text{HF}}$, associated to this Hamiltonian (note that we use $F$ because we work with fixed number of particles; if we worked with fixed chemical potential, we would use the grand canonical potential instead), we first replace the interaction term, $\hat{n}_{x,\uparrow} \hat{n}_{x,\downarrow}$, with the Hartree and Fock terms,
\begin{widetext}
\begin{equation}
\hat{n}_{x,\uparrow} \hat{n}_{x,\downarrow} \rightarrow \hat{n}_{x,\uparrow} \langle \hat{n}_{x,\downarrow} \rangle + \langle \hat{n}_{x,\uparrow} \rangle \hat{n}_{x,\downarrow} - \langle \hat{n}_{x,\uparrow} \rangle \langle \hat{n}_{x,\downarrow} \rangle - c_{x,\uparrow}^{\dagger} c_{x,\downarrow} \langle c^{\dagger}_{x,\downarrow} c_{x,\uparrow} \rangle - \langle c^{\dagger}_{x,\uparrow} c_{x,\downarrow} \rangle c_{x,\downarrow}^{\dagger} c_{x,\uparrow} + \langle c^{\dagger}_{x,\uparrow} c_{x,\downarrow} \rangle \langle c^{\dagger}_{x\downarrow} c_{x,\uparrow} \rangle .
\label{eq-HF_terms}
\end{equation}
%\end{widetext}
Replacing the averages by the mean-field parameters $\vec{m}$ and $n$ (see Appendix A for details), it follows that the mean-field Hamiltonian, $H(\vec m,n)$, corresponding to the Hamiltonian in Eq.~\ref{eq-H_calculateOmega} is
%\begin{widetext}
\begin{equation}
H(\vec m,n) = t \sum\limits_{\langle x,y \rangle,\sigma} c^{\dagger}_{x,\sigma} c_{y,\sigma} + U \sum\limits_{x} \left[ \frac{1}{4} \left( \vec{m}_x^2 - n_x^2 \right) +\frac{1}{2} \left( n_x \hat{n}_x - \vec{m}_x \cdot \hat{ \vec{s}}_x \right) \right] ,
\label{eq-H_MF_nm_implicit}
\end{equation}
\end{widetext}
where $\hat{ \vec{s}}_x$ and $\hat{n}_x$ are the spin and electron density operators at the site $x$, and $\vec{m}_x$ and $n_x$ are the respective mean-field parameters.

The Helmholtz free energy $F(\vec m,n)$ is calculated from $H(\vec m,n)$ using the partition function, $Z(\vec m,n)$,
\begin{equation}
\begin{split}
F(\vec m,n) &= - \frac{1}{\beta} \ln Z(\vec m,n) = - \frac{1}{\beta} \ln \left( \text{Tr} \left( e^{-\beta H(\vec m,n)} \right) \right) \\
&= - \frac{1}{\beta} \text{Tr} \left( \ln \left( 1+e^{-\beta h} \right) \right) + \frac{U}{4} \sum\limits_x \left( \vec{m}_x^2 - n_x^2 \right) ,
\end{split}
\end{equation}
where
\begin{equation}
h_{x \sigma y \sigma'} = t \delta_{\sigma \sigma'} + \frac{U}{2} \left( n_x \delta_{\sigma \sigma'} - \vec{m}_x \cdot \vec{\sigma}_{\sigma \sigma'} \right) \delta_{xy} ,
\end{equation}
Here, $\vec{\sigma}_{\sigma \sigma'}$ is the vector of Pauli matrices.

The important result is that in the Hubbard model, the minimum of the Helmholtz free energy, $F_{\text{HF}}$, corresponds to a saddle point of its mean-field counterpart, $F(\vec m,n)$,
\begin{equation}
F_{\text{HF}} =  \min\limits_{\vec{m}} \max\limits_{n} F(\vec m,n).
\label{eq:maxmin}
\end{equation}
See Appendix A for a detailed discussion of the above relation.

Computing the partial derivatives of $F(\vec m,n)$ with respect to $\vec{m}$ and $n$ and setting them equal to zero, we find the self-consistency relations $n_x = \langle \hat{n}_x \rangle$ and $\vec{m}_x = \langle \hat{ \vec{s}}_x \rangle$. In the case of the Hubbard model, solving the Hartree-Fock equations self-consistently is actually equivalent to finding a saddle point of the mean-field energy, $F(\vec m,n)$. If one fixes the particle density, $n$ on all sites of the lattice, then the mean-field calculation is reduced to finding a minimum of the mean-field energy with respect to the spin density, $\vec m$. Note that these results are for finite temperatures, but they are still valid in the limit $T \rightarrow 0$ in the case of the Hubbard model, as shown in Ref.~\onlinecite{Bach1996}.

References \onlinecite{Langmann2007} and \onlinecite{Woul2007} go on to apply Eq.~\ref{eq:maxmin} to the computation of a mean-field magnetic phase diagram of the  Hubbard model in the square lattice, imposing two restrictions that we do not adopt in this work. Namely, they restrict magnetic phases to ferromagnetic (F), antiferromagnetic (AF), and paramagnetic (P), and additionally impose homogeneous particle density throughout the square lattice, in which case the extremization of the free energy given in Eq.~\ref{eq:maxmin} is reduced to a minimization problem. However, in the case of the Lieb lattice, charge modulation occurs even in the tight-binding limit. For a  mean-field calculation  to yield the correct result in the tight-binding limit, this charge modulation needs to be taken into account. To the extent of our knowledge, this work is the first application of the result in Eq.~\ref{eq:maxmin}  which uses both $n$ and $\vec{m}$ to extremize $F(\vec m,n)$. To accomplish this, we use the generalized HF theory, which turns the original minimization problem for the Helmholtz free energy $F$ into a saddle-point problem for the mean-field Helmholtz free energy $F(\vec m,n)$ (see Appendices).

\section{The Lieb lattice in the tight-binding limit}
\label{section_Lieb_tb}

The Lieb lattice is a square lattice, with a quarter of its atoms removed in a regular pattern. Introducing a different creation operator in each sublattice, $A^\dagger$, $B^\dagger$, and $C^\dagger$, the tight-binding term of the Hamiltonian of such a model, $H_t$, is given by \cite{Nita2013}
\begin{equation}
\begin{array}{l}
t \sum\limits_{x=1}^{L_x} \sum\limits_{y=1}^{L_y} \left[ (A_{x,y}^\dag B_{x,y}   + A_{x,y}^\dag C_{x,y}   + \text{H.c.}) \right. \\
\phantom{aaa} \left. + (A_{x,y}^\dag B_{x,y-1} + A_{x,y}^\dag C_{x-1,y} + \text{H.c.}) \right] .
\end{array}
\label{eq-Hamiltonian_Lieb_tb_complete}
\end{equation}
$L_x$ ($L_y$) is the number of unit cells along the $x$ ($y$) direction. The hopping terms in the first line are intra-unit cell and the remaining are inter-unit cell.  Its eigenvalues originate three energy bands, one of which is flat. The dispersion relation for periodic boundary conditions  is
\begin{equation}
\varepsilon_{\pm} = \pm 2 t \sqrt{\cos^2 \frac{k_x}{2} + \cos^2 \frac{k_y}{2}} ,
\label{eq-Lieb_dispersionrelation}
\end{equation}
for the two non-flat energy bands, where $k_{\alpha} = 2 \pi n_{\alpha} /L_{\alpha}$ with $n_{\alpha} = 0,1,\cdots ,L_{\alpha}$ and $\alpha \in \{ x,y \}$. The flat band is $L_x \times L_y$-fold degenerate with zero energy. These three energy bands are shown in Fig.~\ref{fig-Lieb_tb_dispersionrelation}. The three branches intersect at the point $(k_x,k_y) = (\pi,\pi)$. Expanding the dispersion relation in Eq.~\ref{eq-Lieb_dispersionrelation} around this momentum, we find the Dirac cones $\varepsilon^2 = t^2 (k_x^2 + k_y^2)$. The flat band is built up from B- and C-type orbitals, while the lower and upper bands involve all three lattices A, B, and C. This lack of uniformity in the distribution of the sublattices in the energy bands justifies the difference in the occupation numbers of the sublattices presented in Fig.~\ref{fig-Lieb_tb_E_and_n}.

\section{Interactions and mean-field}

In this section, we add interactions to the tight-binding Hamiltonian of the Lieb lattice, $H_t$, and reduce the quartic dependance of the resulting Hamiltonian on the creation and destruction operators to a quadratic one, using the mean-field approximation \cite{S.Sarker1991}.

The key difference between our approach and previous approaches is that we allow sublattices A, B, and C to have a different average occupation number each, while keeping the total number of particles of the system, $N$, fixed [on each point of the $(n,U)$ phase diagram]. We begin by defining an average particle density on each sublattice,
\begin{equation}
\begin{array}{l}
n_A = n+\delta_A \\
n_B = n+\delta_B \\
n_C = n+\delta_C ,
\end{array}
\end{equation}
along with the number of particles on each sublattice. For sublattice A, this would be $N_A = n_A L$, where $L = L_x L_y$ is the number of unit cells (which in turn equals the number of sites A). To keep the number of particles equal to $N$, we apply the restriction $N_A + N_B + N_C = N$, which is equivalent to
\begin{equation}
\delta_A + \delta_B + \delta_C = 0 .
\label{eq-Nconservation}
\end{equation}
Setting $\delta_A = \delta_B = \delta_C = 0$, we would obtain a lattice with its particles evenly distributed, which is what happens in the usual 2D square lattice: all sites have the same particle density. Aside from total particle number conservation (Eq.~\ref{eq-Nconservation}) and motivated by the symmetry of the lattice, we impose is that $\delta_B = \delta_C$. This gives the important relation $\delta_B = \delta_C = -\delta_A/2 \Rightarrow n_B = \frac{1}{2} (3n-n_A)$, which leaves us with one unknown with respect to which $F(\vec m,n)$ needs to be maximized. We chose to maximize with respect to $\delta_A$.

In our case, we work at $T=0$, so that the mean-field free energy of the system can be found by summing the mean-field energies of the lowest levels that the $N$ particles can occupy (this is the usual Fermi sea). On each point $(n,U)$, the total energy of the system, $E_{\text{HF}}$, is obtained by adding the lowest $N$ eigenvalues of the mean-field Hamiltonian $H_{\text{HF}}$ \cite{Dzierzawa1992,A.Singh1992,Gouveia2014,Gouveia2015},
\begin{equation}
H_{\text{HF}} = \left( \begin{array}{cc}
H_t(\vec{k}) + H_{\delta} & H_m \\
H_m^{\dagger} & H_t(\vec{k} + 2\vec{q}) + H_{\delta}
\end{array} \right) ,
\label{eq-pre_HMF}
\end{equation}
and then adding the diagonal terms
\begin{equation}
\frac{U L}{4} (m_A^2+m_B^2+m_C^2-(n+\delta_A)^2-(n+\delta_B)^2-(n+\delta_C)^2) .
\label{eq-diagterms_HMF}
\end{equation}
The smaller matrices that compose the Hamiltonian $H_{\text{HF}}$ are
\begin{equation}
H_t(\vec{k}) = \left( \begin{array}{cccc}
0 & t (1+e^{i k_y}) & t (1+e^{i k_x}) \\
t (1+e^{-i k_y}) & 0 & 0 \\
t (1+e^{-i k_x}) & 0 & 0
\end{array} \right) ,
\end{equation}
\begin{equation}
H_{\delta} = \frac{U}{2} \left( \begin{array}{cccc}
n+\delta_A & 0 & 0 \\
0 & n+\delta_B & 0 \\
0 & 0 & n+\delta_C
\end{array} \right) ,
\end{equation}
and
\begin{equation}
H_m = -\frac{U}{2} \left( \begin{array}{cccc}
m_A & 0 & 0 \\
0 & m_B e^{i q_y} & 0 \\
0 & 0 & m_C e^{i q_x}
\end{array} \right) .
\end{equation}
The matrix $H_{\text{HF}}$ above is written in the basis $\{ A_{\vec{k}},B_{\vec{k}},C_{\vec{k}},A_{\vec{k}+2\vec{q}},B_{\vec{k}+2\vec{q}},C_{\vec{k}+2\vec{q}} \}$, where the vector $\vec{q} = (q_x,q_y)$ defines the spin orientation in the system, as in the works by Dzierzawa \cite{Dzierzawa1992} and Singh \cite{A.Singh1992}. In this paper, we assume that the spin spiral wavenumber $\vec{q}$ remains the same as in Ref.~\onlinecite{Gouveia2015}, even though we allow the system to have spin and charge modulation. $H_t(\vec{k})$ is the matrix that corresponds to the tight-binding term of the Hamiltonian. All other terms have correspondence with the interaction terms of the Hamiltonian in Eq.~\ref{eq-H_MF_nm_implicit}. Namely,
\begin{equation}
\begin{array}{l}
\frac{U}{4} \sum\limits_x \vec{m}_x^2 \rightarrow \frac{U L}{4} (m_A^2 + m_B^2 + m_C^2) , \\
\\
\frac{U}{4} \sum\limits_x n_x^2 \rightarrow \frac{U L}{4} ((n + \delta_A)^2 + (n + \delta_B)^2 + (n + \delta_C)^2) , \\
\\
\frac{U}{2} \sum\limits_x n_x \hat{n}_x \rightarrow \frac{U}{2} \text{diag} (n+\delta_A , n+\delta_B , n+\delta_C) = H_{\delta} , \\
\\
- \frac{U}{2} \sum\limits_x \vec{m} \cdot \hat{\vec{s}}_x \rightarrow -\frac{U}{2} \text{diag} (m_A , m_B e^{i q_y} , m_C e^{i q_x}) = H_m .
\end{array}
\end{equation}
The extra imaginary coefficients involving components of $\vec{q}$ arise from coupling sites on unit cells other than the cell labelled as $(x,y)$. From this point forward, we consider $t=1$, so that $U$ is given in units of $t$.

\section{Results and discussion}

Our results consist of the values of $m_A$, $m_B$ and $n_A$ which, for each pair $(n,U) \in [0,2] \times [0,20]$, correspond to a saddle point of $E_{\text{HF}}$, the mean-field energy of the Lieb lattice (see Appendix B for a more detailed explanation on how to find these saddle points). From these three quantities, we can calculate $m_C = m_B$, and $n_B = n_C = \frac{1}{2} (3n-n_A)$. We do not impose different occupation or magnetization, we simply let the system choose the values which lead to a saddle point of the mean-field energy. Before performing the calculations for the Lieb lattice, we tested the algorithm for the Hubbard model in a square lattice and found that the occupations of all four sublattices (A, B, C, and D) were the same, i. e., $\delta_A = \delta_B = \delta_C = \delta_D = 0$, while the system chose to have two different magnetizations, $m_A = m_D$ and $m_B = m_C$, reproducing our results in Ref.~\onlinecite{Gouveia2014}.

\begin{figure*}[t!]
\centering
\subfloat[ ]{\label{fig-Lieb_nA}\includegraphics[width=.4 \textwidth]{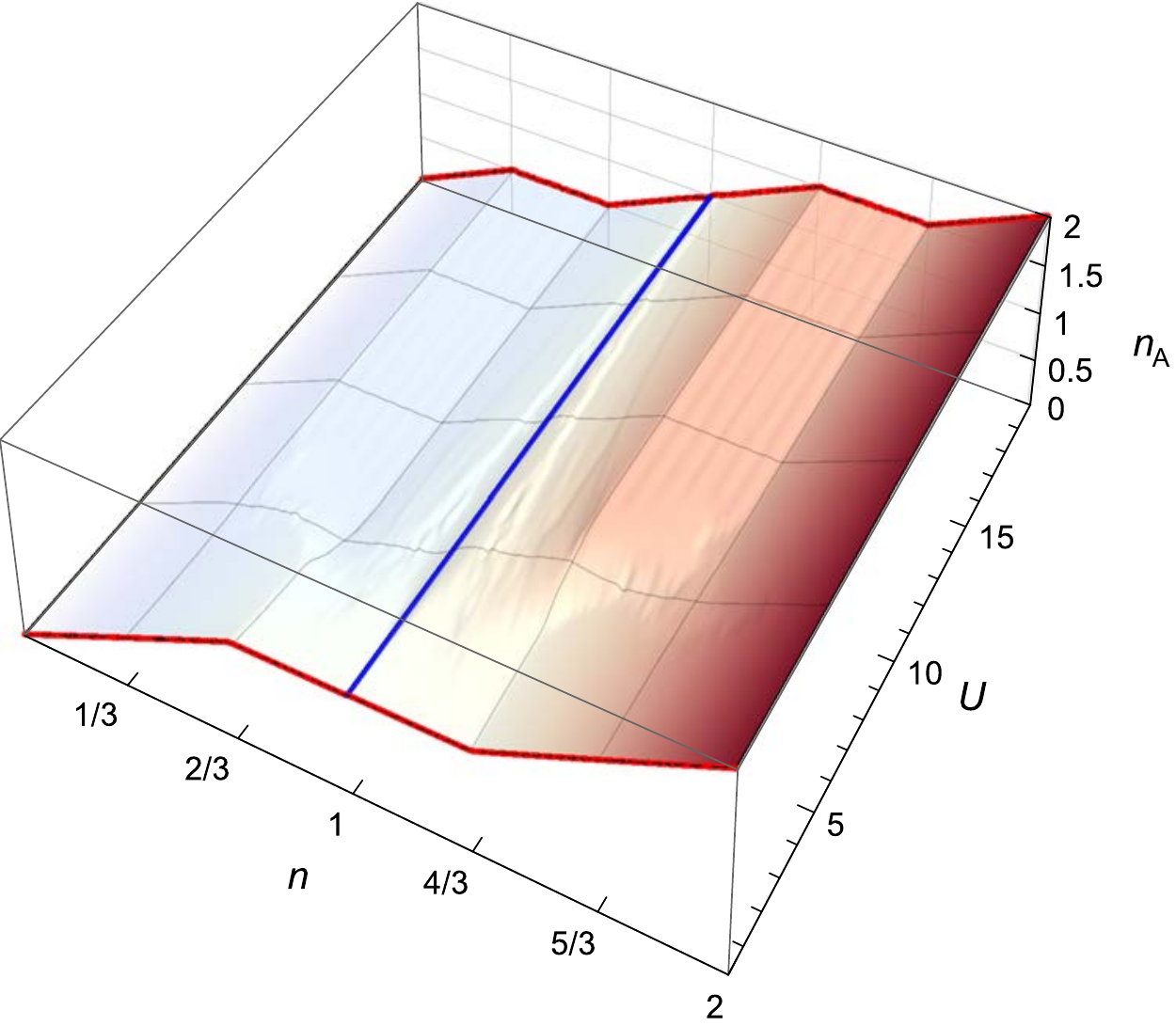}}
\hspace{1cm}
\subfloat[ ]{\label{fig-Lieb_nB}\includegraphics[width=.4 \textwidth]{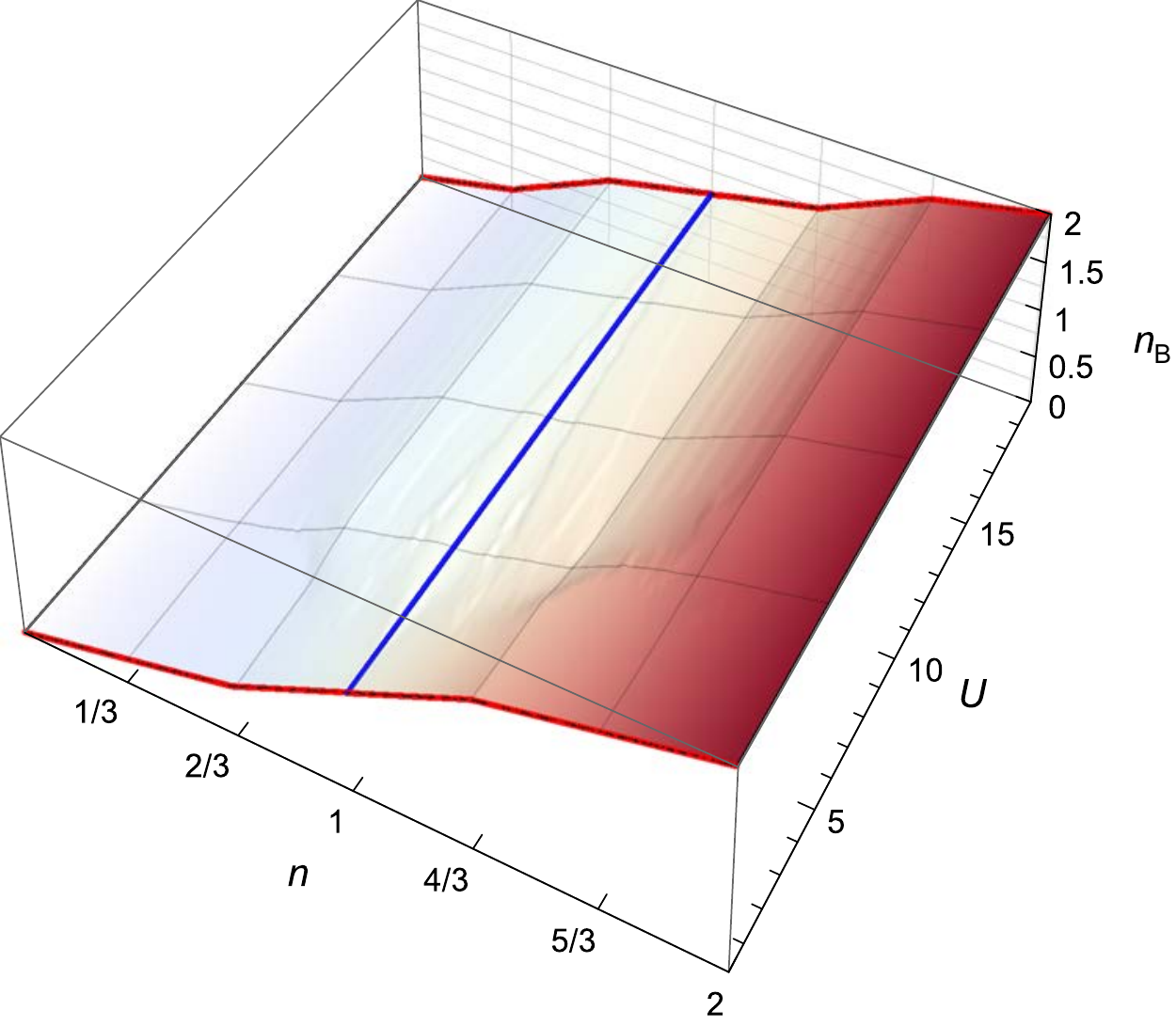}} \\
\subfloat[ ]{\label{fig-Lieb_mA}\includegraphics[width=.4 \textwidth]{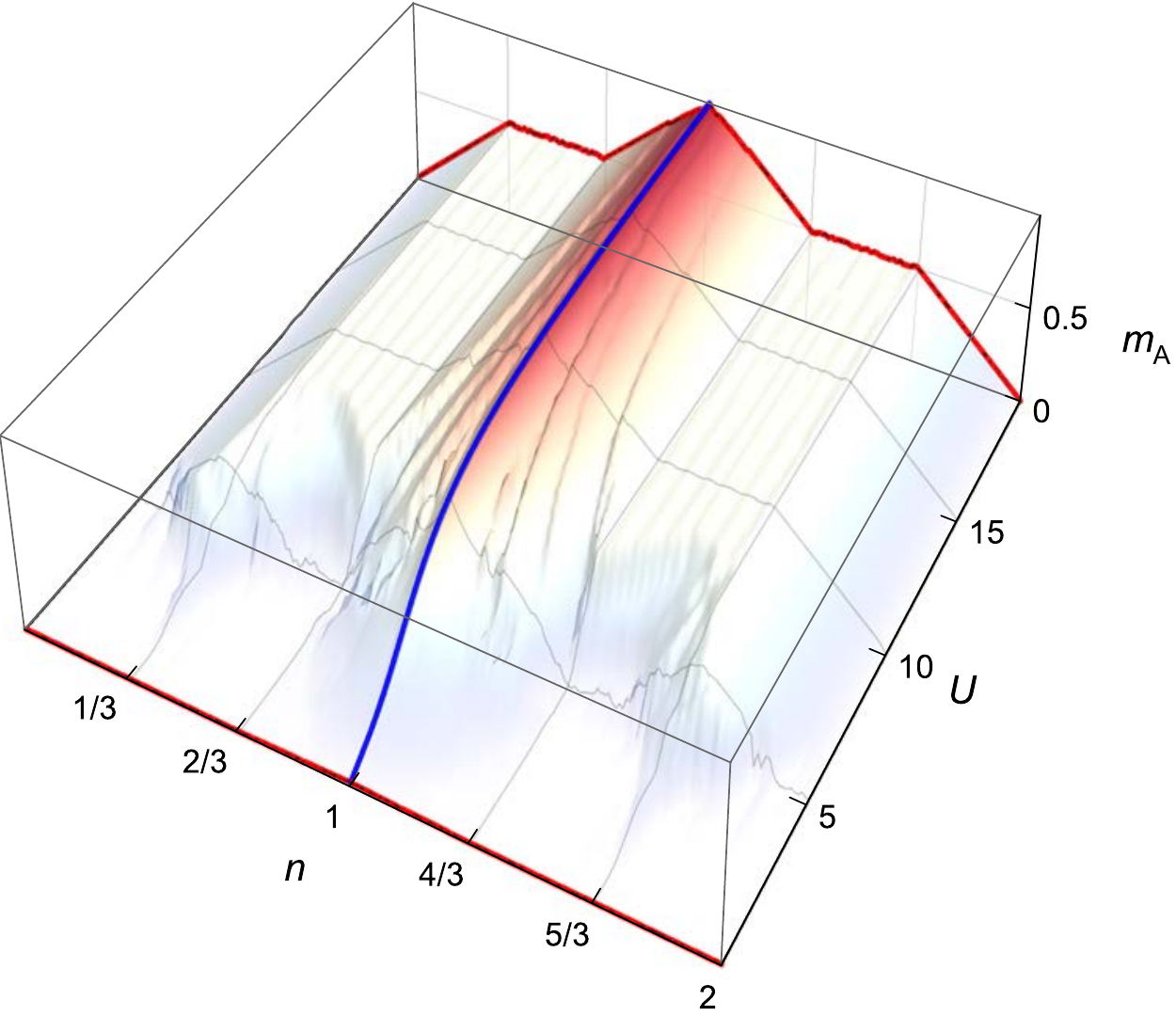}}
\hspace{1cm}
\subfloat[ ]{\label{fig-Lieb_mB}\includegraphics[width=.4 \textwidth]{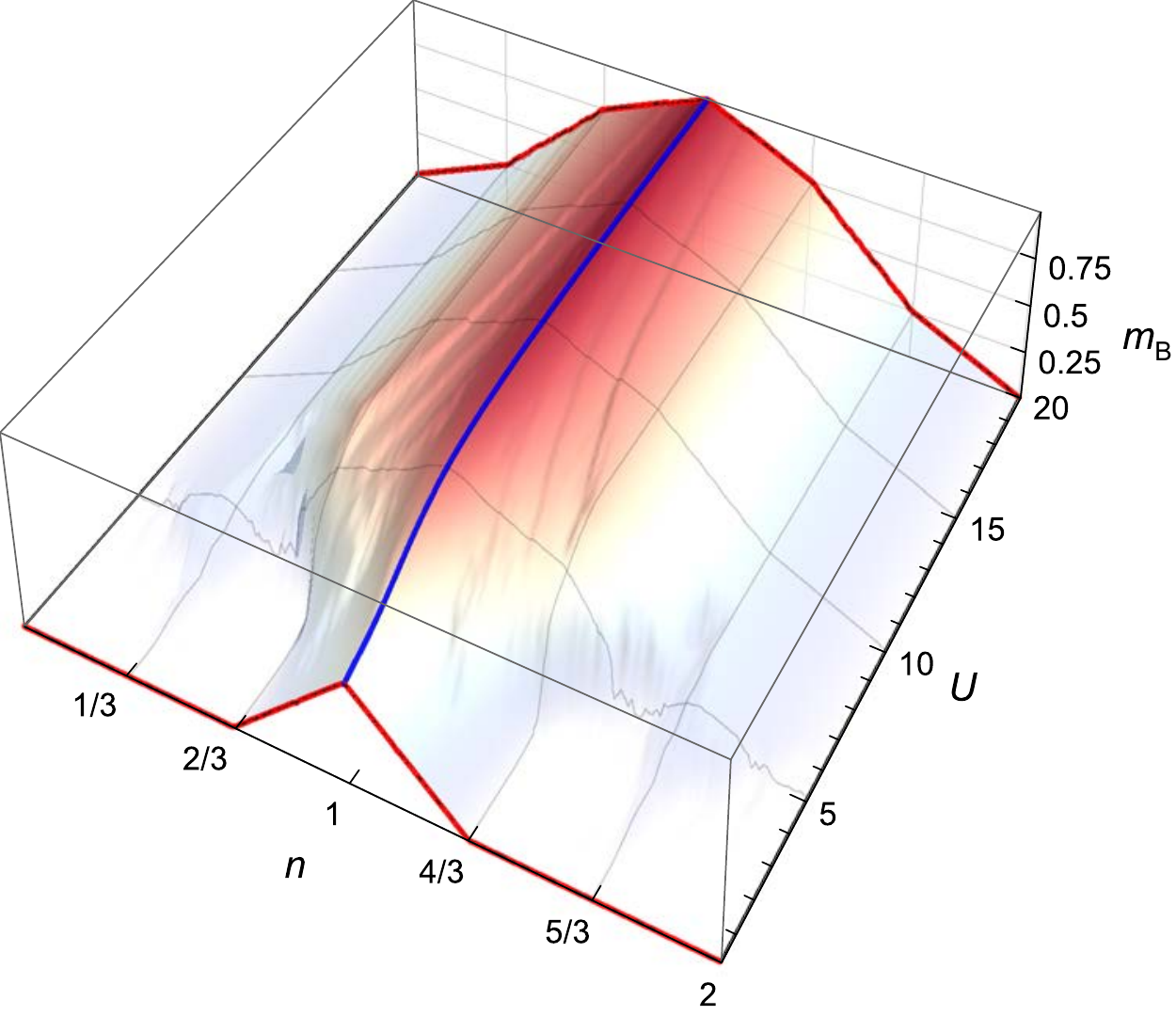}}
\caption{Raw numerical results of our mean-field approach, namely for (a) $n_A$, (b) $n_B$, (c) $m_A$ and (d) $m_B$. The bold red lines highlight the $U=0$ and $U=20$ edges of the plots, and the bold blue line at the center of each plot allows a clearer visualization of each plot at half-filling ($n=1/2$).}
\label{fig-results_Lieb}
\end{figure*}

The results for the Lieb lattice are in Fig.~\ref{fig-results_Lieb}. In the following subsections, we discuss each region of interest in more detail. In particular, we study the low $U$ and high $U$ regions separately, and finally the near-half-filling region ($n \approx 1$).

\subsection{Results near the tight-binding limit ($U \rightarrow 0$)}

As expected, for low $U$, the average occupations of the sublattices, $n_A$ and $n_B = n_C$ (Figs. \ref{fig-Lieb_nA} and \ref{fig-Lieb_nB}, respectively), approach those of the tight-binding limit, described in section \ref{section_Lieb_tb} and plotted in Fig.~\ref{fig-Lieb_tb_E_and_n}. As for the magnetization amplitudes, sublattice A is paramagnetic ($m_A = 0$) for any $n$ and sublattice B displays a behaviour similar to that of $m$ in Ref.~\onlinecite{Gouveia2015} for low $U$. In other words, $m_B$ is finite between $n = 2/3$ and $n = 4/3$ and zero otherwise. As explained below, this difference is due to the fact that the flat band only involves B and C orbitals. A first-order perturbation analysis follows, explaining this result.

Without the $U$ perturbation, the system becomes a Lieb lattice without electron-electron interactions and displays the electronic density in Fig.~\ref{fig-Lieb_tb_E_and_n}. The energy dispersion relation comprises three bands, as in Fig.~\ref{fig-Lieb_tb_dispersionrelation}, which can be doubly occupied with no additional energy cost (because of the absence of $U$). Keep in mind that the dispersive bands comprise A-, B-, and C-type orbitals, while the flat band comprises only B- and C-type orbitals. The introduction of the repulsive first-order perturbation modifies the total energy of the system by shifting the energy bands and adding the diagonal terms of Eq.~\ref{eq-diagterms_HMF}. The dispersive bands are shifted by $+\frac{U \delta_A}{8}$ and the flat band is shifted by $-\frac{U \delta_A}{4}$. Additionally, the flat band (which used to have zero energy) splits into two bands, separated by an amount proportional to $U m_B$.

At zero filling, the energy bands are in their original ($U=0$) position, because all $\delta$ and $m$ are set to zero (having at least one finite $m$ would lead to higher energy due to the term in Eq.~\ref{eq-diagterms_HMF}). As we insert electrons in the system, they occupy the lowest states in the lower dispersive band, with $n_A = 2 n_B = 2 n_C$. Due to the perturbation, this slowly causes the flat band to shift to lower energy and the dispersive bands to shift to higher energy. Note that before the filling $n \approx 2/3$, the system is able to have the lowest energy by displaying paramagnetism ($m_A = m_B = 0$), because, to first order, the only dependance of the energy on $m$ is in the diagonal term of Eq.~\ref{eq-diagterms_HMF}. This dependance is maintained for any $n$, in the case of $m_A$. At $n \approx 2/3$, the lower dispersive band is almost full and, due to the flat bands having shifted by the small amount $-\frac{U \delta_A}{4}$, electrons begin to occupy the flat bands, rather than the dispersive bands. From here on, up to $n=1$, the magnetization amplitude of sublattice B ($m_B$) increases, in order to separate the flat bands into two, and decrease the energy of the lower flat band, which is the one being filled. Due to the high density of states in the flat bands (which are related to B and C atoms only), newly added electrons choose to occupy sublattices B and C, until both flat bands are filled, which occurs at $n \approx 4/3$. Between $n=1$ and $n=4/3$, the magnetization $m_B$ decreases again, because the lower flat band is full and electrons are now occupying the higher flat band, which has energy proportional to $m_B$. For filling $n \in [1,2]$, the behaviour is symmetrical to that of the $n \in [0,1]$ region.

\subsection{Results in the strong coupling limit ($U \gg t$)}
\label{section-results_Ularge}

In this subsection, we discuss our results for high $U$, which we can assume to be nearly identical to those at $U \rightarrow \infty$, for two reasons. Firstly, through inspection of the plots in Fig.~\ref{fig-results_Lieb}, one readily realizes that the behaviour at $U=20$ is approximately the same as, say, $U=15$, and therefore should not change with an increase in $U$. Moreover, if a certain value of $U$ is enough to impose ferromagnetism (and therefore, for $n<1$, the energy bands are singly occupied with spins in the same direction), higher values of $U$ will have no additional effect. Secondly, the analytical results for $U \rightarrow \infty$ which we present in the following paragraphs are in agreement with the numerical results in Fig.~\ref{fig-results_Lieb} for $U=20$.

One important first remark is that, as can be seen from Figs. \ref{fig-Lieb_nA} and \ref{fig-Lieb_nB}, the value of $n_A$ or $n_B$ for $U=20$ along the line $n \in [0,1]$, is the same as for $U=0$ along the line $n \in [0,2]$, but divided by two. This is because, at high $U$, the inequivalence of sublattices is imposed solely by the tight-binding terms of the Hamiltonian (see Eq.~\ref{eq-pre_HMF}). Therefore, the behaviour of $n_A$ and $n_B$ is the same as in the tight-binding limit, albeit with all spins equally aligned. In fact, at $U=0$, the tight-binding bands become doubly occupied without any additional energy cost, while for high $U$ this cost is so high that all tight-binding states (thus, all sublattices) become singly occupied before double occupancies are created.

To study the $U \rightarrow \infty$ limit from a perturbation theory point of view, we begin by setting $t=0$ in the Hamiltonian in Eq.~\ref{eq-pre_HMF} and taking the result as the unperturbed Hamiltonian. Its eigenvalues are
\begin{equation}
\begin{array}{l}
\frac{U}{2} (n_A \pm m_A) , \\
\\
\frac{U}{2} (n_B \pm m_B) , \\
\\
\frac{U}{2} (n_C \pm m_C) .
\end{array}
\label{eq-highU_bands}
\end{equation}
These are six flat bands, with $L$ states each. Positive $m$ and negative $m$ give the same set of eigenvalues, so let us assume positive $m$, with no loss of generality. At $n < 1$, electrons occupy the three lowest energy bands: $\frac{U}{2} (n_A - m_A)$, $\frac{U}{2} (n_B - m_B) $ and $\frac{U}{2} (n_C - m_C) $, so that the total mean-field energy of the system is given by
\begin{widetext}
\begin{equation}
E_{U \rightarrow \infty} = \frac{U L}{4}(m_A^2+m_B^2+m_C^2-n_A^2-n_B^2-n_C^2) + \frac{U}{2} \left[ \sum\limits_{N_A} (n_A-m_A) + \sum\limits_{N_B} (n_B-m_B) + \sum\limits_{N_C} (n_C-m_C) \right],
\end{equation}
\end{widetext}
where we have reintroduced the diagonal terms of Eq.~\ref{eq-diagterms_HMF}. This expression can be simplified using the symmetries mentioned above: (i) $\delta_A + \delta_B + \delta_C = 0$, (ii) $n_B = n_C$ and (iii) $m_B = m_C$. Using these three relations and performing the summations up to some fixed $\widetilde{N}_A = L \widetilde{n}_A$, the total energy for large $U$ and $n<1$ becomes
\begin{widetext}
\begin{equation}
E_{U \rightarrow \infty} = \frac{U L}{4} (m_A^2+2m_B^2-\frac{3}{2} n_A^2-\frac{9}{2} n^2+3 n n_A) + \frac{U L}{2} \left[ \widetilde{n}_A (n_A-m_A) + (3n-\widetilde{n}_A) \left( \frac{1}{2} (3n-n_A)-m_B \right) \right].
\end{equation}
\end{widetext}
We find the ground state energy by taking $\vec{\nabla} E_{U \rightarrow \infty} = \vec{0}$, where the derivatives are taken with respect to $m_A$, $m_B$ and $n_A$. The result is the self-consistency $n_A = \widetilde{n}_A$ and the relations $m_A = n_A$ and $m_B = n_B$. These two relations hold true for $U=20$, as can be realized by comparing $n_A$ with $m_A$, and $n_B$ with $m_B$ in Fig.~\ref{fig-results_Lieb}. Going back to $E_{U \rightarrow \infty}$ and replacing $n_A=m_A$ and $n_B=m_B$, we find that the three bands in Eq.~\ref{eq-highU_bands} become degenerate with zero energy. Note that setting $n_A = m_A$ and $n_B = m_B$ does not lead to a minimum of $E_{U \rightarrow \infty}$, but to a saddle point, as expected. Again, Hartree-Fock mean-field theory is not about finding minima, but rather about finding self-consistency, as explained in section II. Finally, we remark that having $n_A = m_A$ and $n_B = m_B$ is a consequence of a ferromagnetic ground state with the bands being filled with only one spin direction.

The next step in the perturbation analysis is to introduce the hopping terms of the Hamiltonian of Eq.~\ref{eq-pre_HMF}, as the perturbation. This perturbation lifts the degeneracy of the three zero-energy bands. Up to first order, one of them, comprising B- and C-type orbitals, retains zero energy, while each of the other two bands are pushed to positive or negative energy, proportionally to $t$, and comprise orbitals of the three types, A, B, and C. This new energy band configuration mimics that of the tight-binding limit. This justifies the relative filling of the sublattices at high $U$ (Figs. \ref{fig-Lieb_nA} and \ref{fig-Lieb_nB}), and therefore their magnetization $m_A = n_A$ and $m_B = n_B$. We now have a lower band involving A, B, and C orbitals, an intermediate flat band built up from B and C orbitals, and a higher energy band involving all three types of atoms. It follows that, as we begin inserting electrons in the system, sublattice A fills at two times the rate of the B/C sublattices. At $n=1/3$, the lower band (with energy proportional to $-t$) is full and $n_A = 1/4 = 2 n_B = 2 \times 1/8$. Between $n=1/3$ and $n=2/3$, the flat band is filled up using only sublattices B and C, therefore $n_A$ does not change. Finally, for $n \in [2/3,1]$, sublattices A, B, and C become gradually half-filled as the band with energy proportional to $+t$ is filled. For filling $n \in [1,2]$, the behaviour is symmetrical to that of the $n \in [0,1]$ region.

\subsection{Results near half-filling ($n \approx 1$)}

\begin{figure}[t!]
\centering
\subfloat[ ]{\label{fig-2mbma}\includegraphics[width=.4 \textwidth]{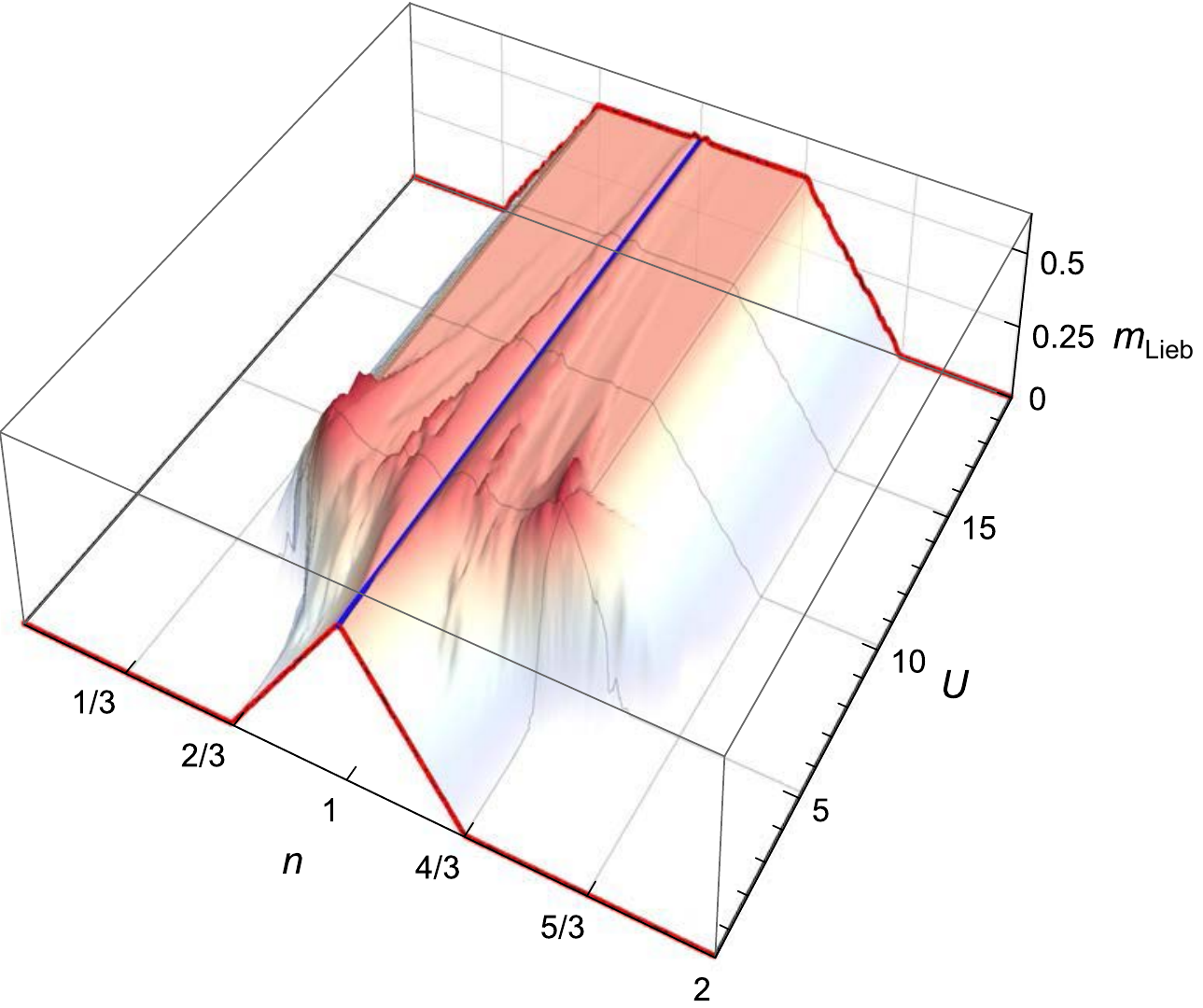}}\\
\subfloat[ ]{\label{fig-phasediagram}\includegraphics[width=.4 \textwidth]{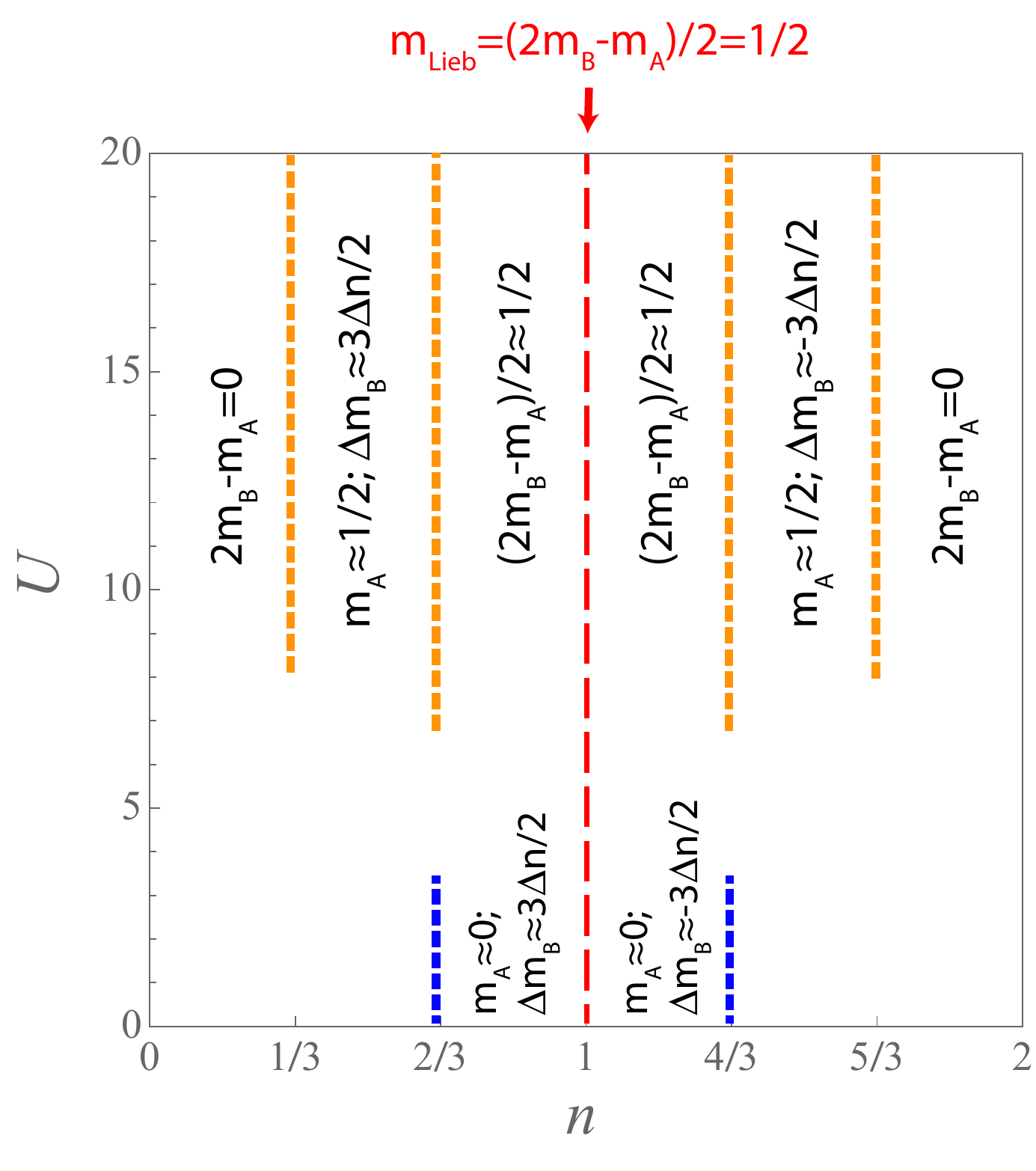}}
\caption{(a) Plot of the  difference in the absolute values of the magnetization of the sublattices, $m_{\text{Lieb}}=\frac{1}{2} (2 m_B - m_A)$, as a function of $n$ and $U$, using the data from the plots in Figs. \ref{fig-Lieb_mA} and \ref{fig-Lieb_mB}. The bold red lines highlight the $U=0$ and $U=20$ edges of the plot, and the bold blue line at the center allows for easier visualization of the behaviour of $\frac{1}{2} (2 m_B - m_A)$ at half-filling. According to a theorem by Lieb \cite{Lieb1989}, the value of $\frac{1}{2} (2 m_B - m_A)$ at half-filling is $1/2$, a value which the plot shows to have been achieved by our mean-field approach. (b) Phase diagram illustrating the relative behavior of the magnetization of sublattices A and B, as a function of $n$ and $U$.}
\label{fig-Lieb_theorem}
\end{figure}

At half-filling, the ground state is a $\vec{q} = (\pi,\pi)$ phase. Lieb's theorem \cite{Lieb1989} states that in bipartite systems, such as our Lieb lattice, the ground state at half-filling is ferrimagnetic \cite{Mielke1993}. In our case, this ferrimagnetic ordering is ferromagnetic within each sublattice, and antiferromagnetic between every two nearest-neighbour sites (see Fig.~\ref{fig-Fsublattice_AFglobal}). Although Lieb's theorem does not provide information on the spin per site, it does state that in our system the quantity $\frac{1}{2} (2 m_B - m_A)$ should be equal to $1/2$. This served as one of the motivations for this work, as previous studies of the Lieb lattice using mean-field \cite{Gouveia2015} failed to yield the value $1/2$ at low $U$. One of the reasons for this was that the magnetization amplitude was the same on all sites, A, B, and C, and the relative occupation of the three sublattices in the tight-binding limit was used for finite $U$.

One can plot the function $\frac{1}{2} (2 m_B - m_A)$, using our results for $m_A$ and $m_B$, in Figs. \ref{fig-Lieb_mA} and \ref{fig-Lieb_mB}. Such a plot can be found in Fig.~\ref{fig-2mbma}.  At exactly $n=1$ (the bold blue line at the center of the plot), our generalized Hartree-Fock approach succeeds in yielding the result $\frac{1}{2} (2 m_B - m_A) = \frac{1}{2}$, thus verifying Lieb's theorem. This leads to the conclusion that our mean-field study, allowing for modulation of $m$ and $n$ in the Lieb lattice, produces more accurate results than imposing the same magnetization and electronic density in the whole lattice. Moreover, this plot indicates that  Lieb's theorem is verified approximately in the wider $n \in [2/3,4/3]$ range.

The phase diagram given by Fig.~\ref{fig-phasediagram} was constructed using the relative behavior of the magnetization of sublattices A and B. The dashed lines indicate the boundaries where the behavior of the magnetization of each sublattice changes. The magnetization $m_{\text{Lieb}}$ is exactly $1/2 $ at the central (red) dashed line ($n=1$). For small $U$, within the interval  $2/3<n<4/3$, only sublattice B has finite magnetization. For large $U$, in the  $n \in [2/3,4/3]$ range,  Lieb's theorem is verified approximately. Note that this range includes not only the ferromagnetic region of the phase diagram of Fig.~\ref{fig-diagram_standard_Lieb}a, but surprisingly also the adjacent spiral regions.

One other theorem, by Lieb and collaborators, refers to the particle density in the sublattices \cite{Lieb1993a}. This theorem states that in the bipartite Hubbard model, there are no charge density modulations at half-filling. In other words, it means that at half-filling all sublattices are equally occupied and therefore half-filled, $n_A = n_B = n_C = 1$. Plotting $n_A$ and $n_B$ as a function of $U$ at fixed $n=1$, using our results for $n_A$ and $n_B$ (shown in Figs. \ref{fig-Lieb_nA} and \ref{fig-Lieb_nB}, respectively), we checked that we obtained the lines $n_A = 1$ and $n_B = 1$. As discussed in Section \ref{section-results_Ularge} of this paper, the behaviour of $n_A$ and $n_B$ at large $U$ consists of two compacted copies of the behaviour at zero $U$. This holds true for any bipartite system. Consequently, for large $U$, not only are there not charge density modulations at half-filling, but also there are no charge modulations at $n=1/2$ and $n=3/2$.

The contrasting behavior in Fig.\ref{fig-phasediagram} for large and small $U$ can be interpreted as a consequence of the modifications of the mean-field energy dispersion  as $U$ increases. More precisely, for small $U$,  one has a single quasi-flat band (twice degenerate with energy $\varepsilon \sim 0$), while for large $U$, one has two non-degenerate one-particle quasi-flat bands well separated in energy ($\varepsilon \sim 0$ and $\varepsilon \sim U$). Note that the flat bands for large $U$  are present in the exact solution of the Hubbard model in the subspace of eigenstates associated with saturated ferromagnetism. As one can conclude from Fig.~\ref{fig-2mbma}, the difference in the absolute values of the magnetizations of the sublattices grows only when a flat band is being filled and in fact, we can write
\begin{equation}
m_{\text{Lieb}}=\dfrac{1}{2}(2m_B-m_A)\approx  \text{filling of flat bands},
\end{equation}
\emph{for filling $n\leq 1$, both for large and small $U$.} For $n >1$, one has the reflected behavior of $n\leq 1$.
\section{Conclusions}

In summary, we have studied the Lieb lattice using a mean-field approach and allowing for charge and spin density modulation. Although theory about the correspondence between Hartree-Fock self-consistency and saddle points of the mean-field energy is relatively old (20 years old), to the extent of our knowledge, this is the first time it is in fact applied to a system where charge  modulation is known to occur. We have found that, in the limits of low interaction ($U \rightarrow 0$) and very high interaction ($U \rightarrow \infty$) results agree with what one would expect. Namely, the relative occupation of sublattices A and B of the bipartite Lieb lattice (where sublattice A comprises the atoms with four nearest neighbours, and B denotes the remaining atoms) in the tight-binding limit coincides with the results in the literature (for instance, in Ref.~\onlinecite{Wang2014}). We have also found that the profile of the relative occupation of the sublattices in both strong-coupling and tight-binding are analogous, and one can be inferred if the other is known. The argument is relatively simple. On the one hand, in the $U=0$ limit, the energy dispersion is governed by the tight-binding terms of the Hamiltonian only and the energy bands in the case of the Lieb lattice are as depicted in Fig.~\ref{fig-Lieb_tb_dispersionrelation}. On the other hand, in the limit $U \gg t$, the energy bands are separated by a very high energy gap (of the order of $U$), the lower bands corresponding to no double occupancies and the higher bands to double occupancies. The gradual filling of the system is done by singly filling all sites and only then jumping to the higher  bands and doubly occupying all sites. Each one of these two filling regimes follows the relative occupation of sublattices that occurs in the tight-binding limit.

At half-filling, our numerical mean-field results verify two important exact results for the Hubbard model. Firstly, one theorem \cite{Lieb1989} states that in bipartite lattices with more B-type atoms than A-type atoms, the total spin per unit cell at half-filling is equal to $m_{\text{Lieb}} = (|B| - |A|)/2$, where $|x|$ denotes the number of $x$-type atoms a unit cell. In the case of our bipartite Lieb lattice, we get $(2-1)/2 = 1/2$. Our results are plotted in Fig.~\ref{fig-Lieb_theorem}. At exactly half-filling we obtained the expected value $1/2$. Additionally, for large $U$, in the $n \in [2/3,4/3]$ range, we found that $m_{\text{Lieb}} \approx 1/2$. Interestingly, this range includes not only the ferromagnetic region of the phase diagram of Fig.~\ref{fig-diagram_standard_Lieb}a, but also the adjacent spiral regions. Secondly, another theorem \cite{Lieb1993a} states that in the bipartite Hubbard model, there are no charge density modulations at half-filling, that is, at half-filling all sublattices are equally occupied and half-filled. Our numerically calculated relative occupations of sublattices A and B of a bipartite Lieb lattice, shown in Figs. \ref{fig-Lieb_nA} and \ref{fig-Lieb_nB}, are in agreement with this theorem. In addition, we found that, for large $U$, not only are there no charge density modulations at half-filling, but also there are no charge modulations at $n=1/2$ and $n=3/2$.

Away from half-filling, we found that, for large and small $U$, the difference in the absolute values of the sublattice magnetizations ($m_{\text{Lieb}}$) grows or decreases only when a flat band is being filled and furthermore, for $n \leq 1$, $m_{\text{Lieb}}$ is given approximately by the filling of the flat bands. Note that $m_{\text{Lieb}}$ is the unit cell magnetization in the case of the $\vec{q}=(\pi,\pi)$, the ferrimagnetic phase of the Lieb lattice found at half-filling.

We suggest that much of the above discussion is valid in the case of other bipartite lattices with flat bands in the energy dispersion. In the case of non-bipartite lattices with more than two types of atoms in the unit cell, the analysis is more complex, because, for instance, an antiferromagnet configuration between sublattices may not be commensurate with the unit cell. Bipartite lattices of various geometries can be realized  by manipulating quantum dot arrays \cite{Tamura2000} or cold atoms in optical lattices \cite{Goldman2011}.

\section*{Appendix A: Min-Max theorem for the Hubbard model}

In this appendix, we discuss how the mean-field method should be applied to the Hubbard model if particle density is not fixed and in particular, we justify Eq.~\ref{eq:maxmin}. Our approach follows the method presented in  Ref.~\onlinecite{Bach1996}. We start by presenting the usual mean-field approach and then we explain, using the method presented in Ref.~\onlinecite{Bach1996}, why this approach is somewhat misleading.

\subsection{The usual mean-field method}
We begin with the Hubbard Hamiltonian, given by
\begin{equation}
H = t \sum\limits_{\langle x,y \rangle , \sigma} c_{x,\sigma}^{\dagger} c_{y,\sigma} + U \sum\limits_{x} \hat{n}_{x,\uparrow} \hat{n}_{x,\downarrow}
\label{eq-H_calculateOmega_appendix}
\end{equation}

We then replace the interaction term with the Hartree and Fock terms (see Eq.~\ref{eq-HF_terms}). These terms are obtained by considering that each fermionic operator only deviates slightly from its mean value, so that in a product of operators, we can neglect quadratic terms in these deviations.

The operators $c_{x,\uparrow}^{\dagger} c_{x,\uparrow}$ and $c_{x,\uparrow}^{\dagger} c_{x,\downarrow}$ can be identified with the particle density operator $n_x$ and the $s_x^+$ operator. For consistence with the notation in reference \cite{Woul2007}, the spin operators in this appendix are
\begin{equation}
\begin{array}{l}
c_{\uparrow}^{\dagger} c_{\downarrow} = s^+ = \frac{1}{2} (s^x + i s^y) \\
c_{\downarrow}^{\dagger} c_{\uparrow} = s^- = \frac{1}{2} (s^x + i s^y) \\
c_{\uparrow}^{\dagger} c_{\uparrow} - c_{\downarrow}^{\dagger} c_{\downarrow} = s^z .
\end{array}
\label{eq-spin_operators_Woul}
\end{equation}
The vector $\hat{\vec{s}}_x$ is the spin density operator on site $x$. The peculiarity of this definition of $s^+$ and $s^-$ is the factor $1/2$ which is often included in operators $s^x$ and $s^y$ instead. Replacing this in the interaction term of the Hamiltonian in Eq.~\ref{eq-H_calculateOmega_appendix} gives
\begin{widetext}
\begin{equation}
H_{\text{HF}} = t \sum\limits_{\langle x,y \rangle,\sigma} c^{\dagger}_{x,\sigma} c_{y,\sigma} + U \sum\limits_{x} \left[ \frac{1}{4} \left( \langle \hat{ \vec{s}}_x \rangle^2 - \langle \hat{n}_x \rangle^2 \right) +\frac{1}{2} \left( \langle \hat{n}_x \rangle \hat{n}_x - \langle \hat{ \vec{s}}_x \rangle \cdot \hat{ \vec{s}}_x \right) \right]  ,
\label{eq-H_MF_nm_implicit_appendix}
\end{equation}
We now replace the averages by the mean-field parameters $\vec{m}_x$ and $n_x$. These parameters can be identified with the mean values of the magnetization and particle density, respectively, upon extremization of the mean-field free energy, i.e., when the self-consistency equations are satisfied. The Hartree-Fock Hamiltonian becomes
\begin{equation}
H(\vec m,n) = \sum\limits_{x,y,\sigma,\sigma'} h_{x\sigma y\sigma'} c_{x,\sigma}^{\dagger} c_{y,\sigma'} + \frac{U}{4} \sum\limits_x \left( \vec{m}_x^2 - n_x^2 \right) ,
\end{equation}
where
\begin{equation}
h_{x \sigma y \sigma'} = t \delta_{\sigma \sigma'}  + \frac{U}{2} \left( n_x \delta_{\sigma \sigma'} - \vec{m}_x \cdot \vec{\sigma}_{\sigma \sigma'} \right) \delta_{xy} .
\label{eq-h_supermatrix}
\end{equation}
\end{widetext}
Here, $\vec{\sigma}_{\sigma \sigma'}$ is the vector of Pauli matrices.
The function $F(\vec m,n)$ is calculated from $H(\vec m,n)$ using the partition function,
\begin{widetext}
\begin{equation}
\begin{array}{l}
F(\vec{m}, n) = - \frac{1}{\beta} \ln Z(\vec {m},n) = - \frac{1}{\beta} \ln \left( \text{Tr} \left( e^{-\beta H(\vec m,n)} \right) \right) \\
\\
\phantom{aaa} = - \frac{1}{\beta} \text{Tr} \left( \ln \left( 1+e^{-\beta h} \right) \right) + \frac{U}{4} \sum\limits_x \left( \vec{m}_x^2 - n_x^2 \right) .
\end{array}
\end{equation}
\end{widetext}

At this stage, one usually finds the minimum free energy, $F_{\text{HF}}$, by minimizing $F(\vec{m}, n) $ with respect to the mean-field parameters and this would lead to the usual self-consistency equations \cite{Bruus2004}. Clearly, this works for fixed particle density, but in the previous expression we allow for variable particle density and the respective quadratic term has a negative coefficient. If one imposed a minimization with respect to the parameter $n_x$, convergence would not be achieved in a numerical approach, unless one limits  the possible values of $n_x$ to a certain interval, in which case the result of the numerical minimization would lie at the boundary of this interval. This reflects the fact that one should not minimize with respect to the parameter $n_x$, but instead maximize, as we explain in the next subsection.

\subsection{A different perspective for the mean-field method}
In this subsection, we present the mean-field approach which should be applied when one takes into account the possibility of non-uniform particle density in a lattice. For a rigorous proof see, for instance, Refs. \onlinecite{Woul2007,Bach1994,Bach1996}.

Our goal is to know the thermal equilibrium state of the system which minimizes the free energy. These states are defined in terms of density matrices. Having an exact free energy would require having an exact partition function, which in turn would require an exact diagonalization of the Hubbard model. The Hartree-Fock method provides an approximation to the exact equilibrium state, in terms of many-body states of non-interacting particles, replacing the quartic terms of the Hamiltonian by one-particle potentials (which are adjusted to provide the best possible approximation). The free energy obtained in the Hartree-Fock method provides an upper bound to the exact free energy (a consequence of the variational theorem). Since the particles are independent, the Hartree-Fock state can be written as a one-particle density matrix, $\gamma_{ij}=\langle c_i^\dagger c_j \rangle$ \cite{Bach1996}. The objective of the  mean-field method is indeed to find the minimum free energy in the set of free energies associated with the possible states of $N$ independent particles (or equivalently, associated with the possible one-particle density matrices),
\begin{equation}
F_{HF} = \min\limits_{\gamma} F ( \gamma ) = \min\limits_{\gamma} [E( \gamma )- \frac{1}{\beta}S(\gamma)] .
\end{equation}
where $\beta$ is the einverse temperature.
In the previous expression, no mean field parameters are present and the free energy is determined for each $\gamma$ using the exact Hamiltonian (the mean field approximation is associated with the state, not with the Hamiltonian). However, calculating $\min_{\gamma} F ( \gamma )$ going through all $\gamma$ is not pratical, so one introduces mean-field parameters. In the next paragraphs, we introduce these parameters, following a derivation in Ref.~\onlinecite{Bach1996}. For now, let us assume zero temperature.

In the case of the Hubbard Hamiltonian, the Hartree-Fock energy functional, $E( \gamma )$, can be written as (Ref.~\onlinecite{Bach1996} Lemma 2)
\begin{equation}
E( \gamma ) = \text{Tr} [ T \gamma ] + U \sum\limits_{x} [\langle \hat{n}_x \rangle^2 -  \langle \hat{\vec{s}}_x  \rangle^{2}] ,
\label{eq-energy_gamma}
\end{equation}
where $x$ labels the lattice sites and $T$ is the matrix whose elements $t_{xy}$ are the transition amplitudes of an electron to move from site $x$ to site $y$ or vice versa. The averages of the electronic and spin densities at site $x$ are
\begin{equation}
\begin{array}{c}
\langle \hat{n}_x \rangle = \text{Tr} \left[ \hat{\mathbb{I}} \cdot \gamma \right] , \\
\langle \hat{\vec{s}}_x  \rangle = \text{Tr} \left[ ( \hat{\mathbb{I}} \otimes \hat{ \vec{\sigma}} ) \cdot \gamma \right] ,
\end{array}
\end{equation}
respectively.

The first step to introduce the variational parameters is to use the simple fact that
\begin{equation}
x^2 \geq 2xy-y^2 \hspace{4 mm} \text{for all $x,y \in \mathbb{R}^n$},
\end{equation}
where equality holds if and only if $x=y$. This implies
\begin{equation}
x^2 = \max\limits_{y} (2xy-y^2) ,
\end{equation}
and therefore, we can write
\begin{equation}
\langle \hat{n}_x \rangle^2 = \max\limits_{n_x} \left\{ 2\langle \hat{n}_x \rangle n_x - n_x^2 \right\}
\label{eq:d0}
\end{equation}
and
\begin{equation}
-\langle \hat{\vec{s}}_x  \rangle^2 = \min\limits_{\vec{m}_x} \left\{ \vec{m}_x^2 - 2 \langle \hat{\vec{s}}_x  \rangle \cdot \vec{m}_x \right\} .
\label{eq:vecd}
\end{equation}
where $n_x$ and $\vec{m}_x$ are, at this point, an arbitrary constant and vector. Inserting this into Eq.~\ref{eq-energy_gamma}, the energy for a given state $\gamma$ at zero temperature assumes the form
\begin{equation}
E (\gamma) = \min\limits_{\vec{m}} \max\limits_{n} \left\{E(n , \vec{m} , \gamma ) \right\} ,
\end{equation}
where $E(n , \vec{m} , \gamma )$ is given by
\begin{widetext}
\begin{equation}
E(n , \vec{m} , \gamma ) = \text{Tr} \left[ \left(T - U\sum\limits_{\text{all sites}} \left[ n_x \cdot \mathbb{ \hat I} - \vec{m}_x \cdot \hat{ \vec{\sigma}} \right]\right) \gamma \right]+ U \sum\limits_{\text{all sites}} \left( \vec{m}_x^2 - n_x^2 \right) ,
\label{eq:energyvecdd0gamma}
\end{equation}
\end{widetext}

The aim of the mean-field method is to obtain
\begin{equation}
\min\limits_{\gamma}  E (\gamma) = \min\limits_{\gamma}  \min\limits_{\vec{m} } \max\limits_{n } \left\{E(n , \vec{m} , \gamma ) \right\} .
\end{equation}
The crucial point of the mean-field method is the possibility of exchanging the order of the extremization, which is implicitly done in standard mean-field. In fact, if particle density is fixed, one has
\begin{equation}
\min\limits_{\gamma}  E (\gamma)
= \min\limits_{\gamma}  \min\limits_{\vec{m} } \left\{E( \vec{m} , \gamma ) \right\}
= \min\limits_{\vec{m} }  \min\limits_{\gamma} \left\{E( \vec{m} , \gamma ) \right\} ,
\end{equation}
since the order of minimization is irrelevant, and one recovers the usual mean-field picture, in which one has to minimize the energy with respect to the mean-field parameter $\vec{m}$.

On the other hand, if the particle density $n_x$ is allowed to actually depend on $x$, in Ref.~\onlinecite{Bach1996} it was shown that indeed
\begin{equation}
\min\limits_{\gamma}  \min\limits_{\vec{m} } \max\limits_{n } \left\{E(n , \vec{m} , \gamma ) \right\}
=
\min\limits_{\vec{m} } \max\limits_{n } \min\limits_{\gamma}  \left\{E(n , \vec{m} , \gamma ) \right\} .
\label{eq:minmaxmaxmim}
\end{equation}
Here, we present a simple physical interpretation for this result.

First, one should note that Eq.~\ref{eq:energyvecdd0gamma} for given $\vec{m}$ and $n$ corresponds to the energy of independent particles subjected to one-particle potentials which depend on the given $\vec{m}$ and $n$. Looking at the left-hand side of Eq.~\ref{eq:minmaxmaxmim}, after one has  minimized and maximized $E(n , \vec{m} , \gamma )$ with respect to $\vec{m}$ and $n$ respectively, the remaining minimization (with respect to $\gamma$) will lead, at zero temperature, to a filled Fermi sea associated to these one-particle potentials. These one-particle potentials might, for instance, generate imbalance between the number of up and down spins and lead to finite magnetization.

Second, Eqs. \ref{eq:d0} and \ref{eq:vecd}, associated with minimization and maximization with respect to $\vec{m}$ and $n$, imply the self-consistency equations
\begin{equation}
\begin{array}{c}
\langle \hat{n}_x \rangle = n_x, \\
\langle \hat{\vec{s}}_x  \rangle= \vec{m}_x ,
\end{array}
\end{equation}
and therefore, computing the left-hand side of Eq.~\ref{eq:minmaxmaxmim} can be interpreted as the following instruction: "Among the set of one-particle density matrices that satisfy the self-consistency equations, find the one with minimum energy". As we said above, this minimum energy corresponds to a filled Fermi sea at zero temperature, so in the set of states corresponding to filled Fermi seas, there is one that satisfies the self-consistency equations. This helps us understand the right-hand side of Eq.~\ref{eq:minmaxmaxmim} in the following way: "Among the set of all filled Fermi seas, find (the energy of) the state which satisfies the self-consistency equations".

This argument can be generalized for finite temperature \cite{Bach1996}, taking into account the entropy contribution. In this case, instead of  filled Fermi seas, one has the one-particle density matrices that minimize the free energy in the case of independent particles. If instead of fixed number of particles, one imposes a fixed chemical potential (in which case the grand-canonical potencial would replace the free energy), this one-particle density matrix would become the Fermi-Dirac distribution function.

\section*{Appendix B: How to extremize $E_{\text{HF}}$}

In this appendix, we explain in more detail the algorithm we used for finding saddle points.

Our first approach to extremize $E_{\text{HF}}$ was perhaps the most intuitive non-brute force one. It consisted in maximizing $E_{\text{HF}}$ with respect to $\delta_A$, $\delta_B$, and $\delta_C$ using our results for $q_x$, $q_y$, $m_A$ and $m_B$ of the Hubbard model in a square lattice \cite{Gouveia2014}. The lattice size was also kept at $100 \times 100$. Due to the two restrictions imposed (fixed number of particles in the system and equivalence of B and C sites), we are left with one parameter as maximizer of $E_{\text{HF}}$. We used $\delta_A$. We note that the minimum of $E_{\text{HF}}$ with respect to $\delta_A$ was found to be negative infinity (using the results in Ref.~\onlinecite{Gouveia2014} for $m_A$ and $m_B$ as starting points). Fixing $q_x$, $q_y$, $m_A$ and $m_B$ means that the magnetic phases are kept the same as our previous ones, only the occupation numbers in the sublattices change. However, finding saddle points by starting from a minimum with respect to one direction (the $m$ direction) and then finding the maximum with respect to an orthogonal direction (the $n$ direction) turned out to diverge on most points of the phase diagram. In fact, the suggestion that saddle points of a function can be found by starting at a random point in the function, minimizing the function with respect to the minimizer variables, and then using those new points to maximize the function with respect to the maximizer variables, is false. As a matter of fact, this statement remains false even in a more generalized case. One might think that by successively minimizing and maximizing the function, one would eventually reach a saddle point. This is also not necessarily true. When trying to use this method to extremize $E_{\text{HF}}$, one finds that, in most points of the diagram, the value of $E_{\text{HF}}$ resulting from the last maximization and the value of $E_{\text{HF}}$ resulting from the last minimization differ by orders of magnitude comparable to those of $E_{\text{HF}}$ itself. Additionally, the values of $m_A$, $m_B$, $\delta_A$ and $\delta_D$ which extremize $E_{\text{HF}}$ (or so one thought) do not converge on each successive iteration, they alternate between several possible results.

\begin{figure*}[t!]
\centering
\subfloat[ ]{\label{fig-saddle_test_0}\includegraphics[width=.33 \textwidth]{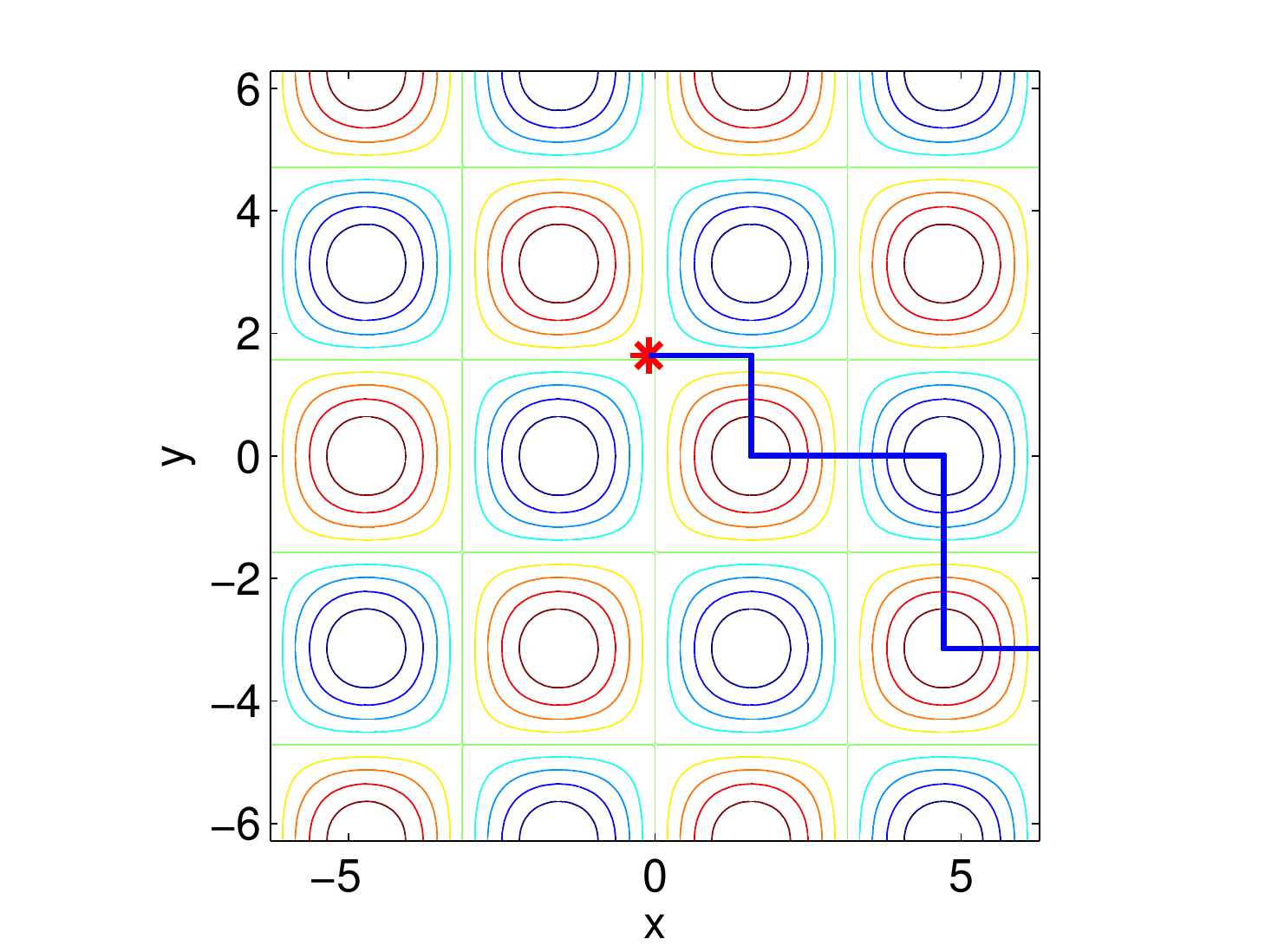}}
\subfloat[ ]{\label{fig-saddle_test_pi20}\includegraphics[width=.33 \textwidth]{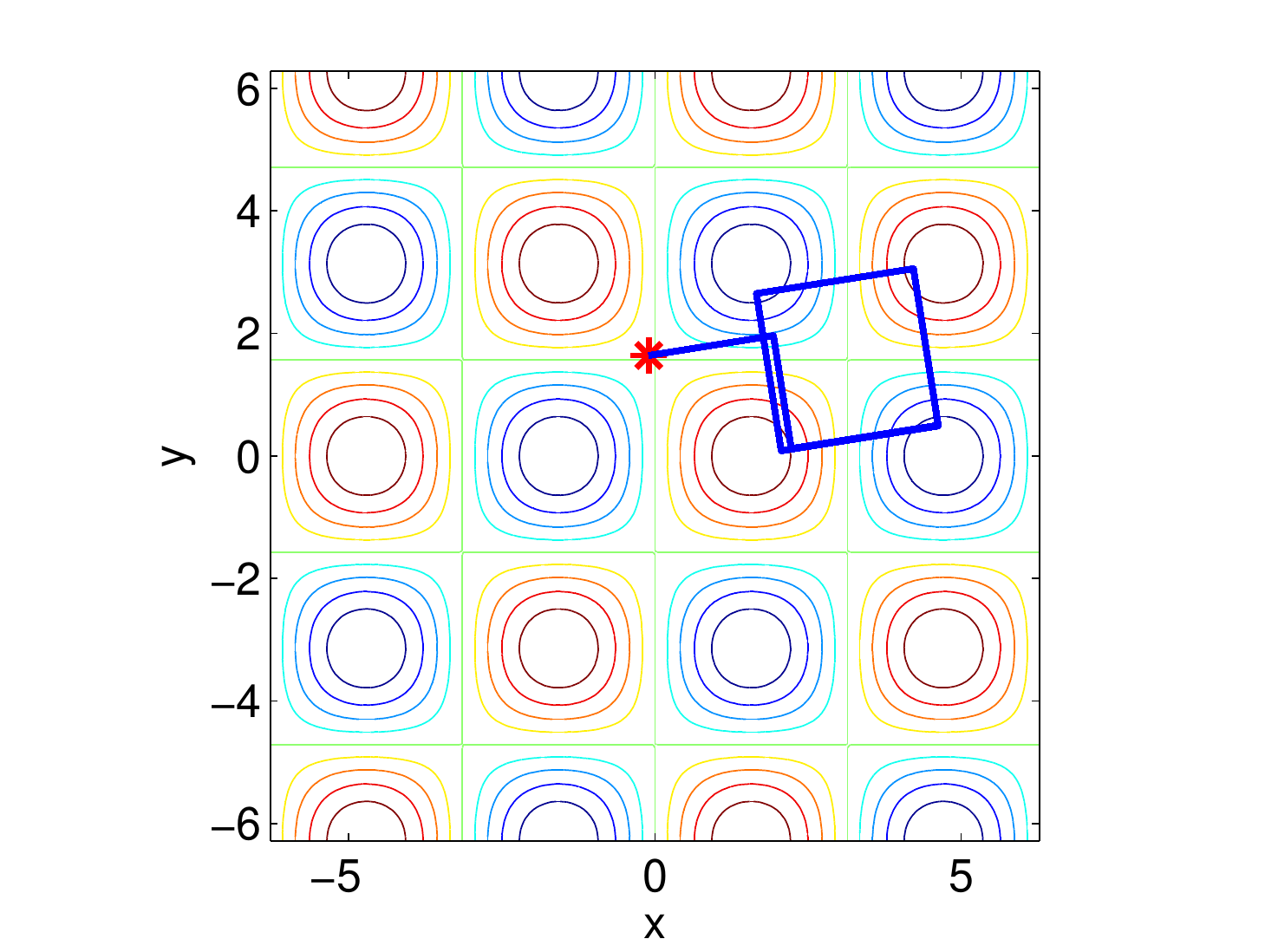}}
\subfloat[ ]{\label{fig-saddle_test_success}\includegraphics[width=.33 \textwidth]{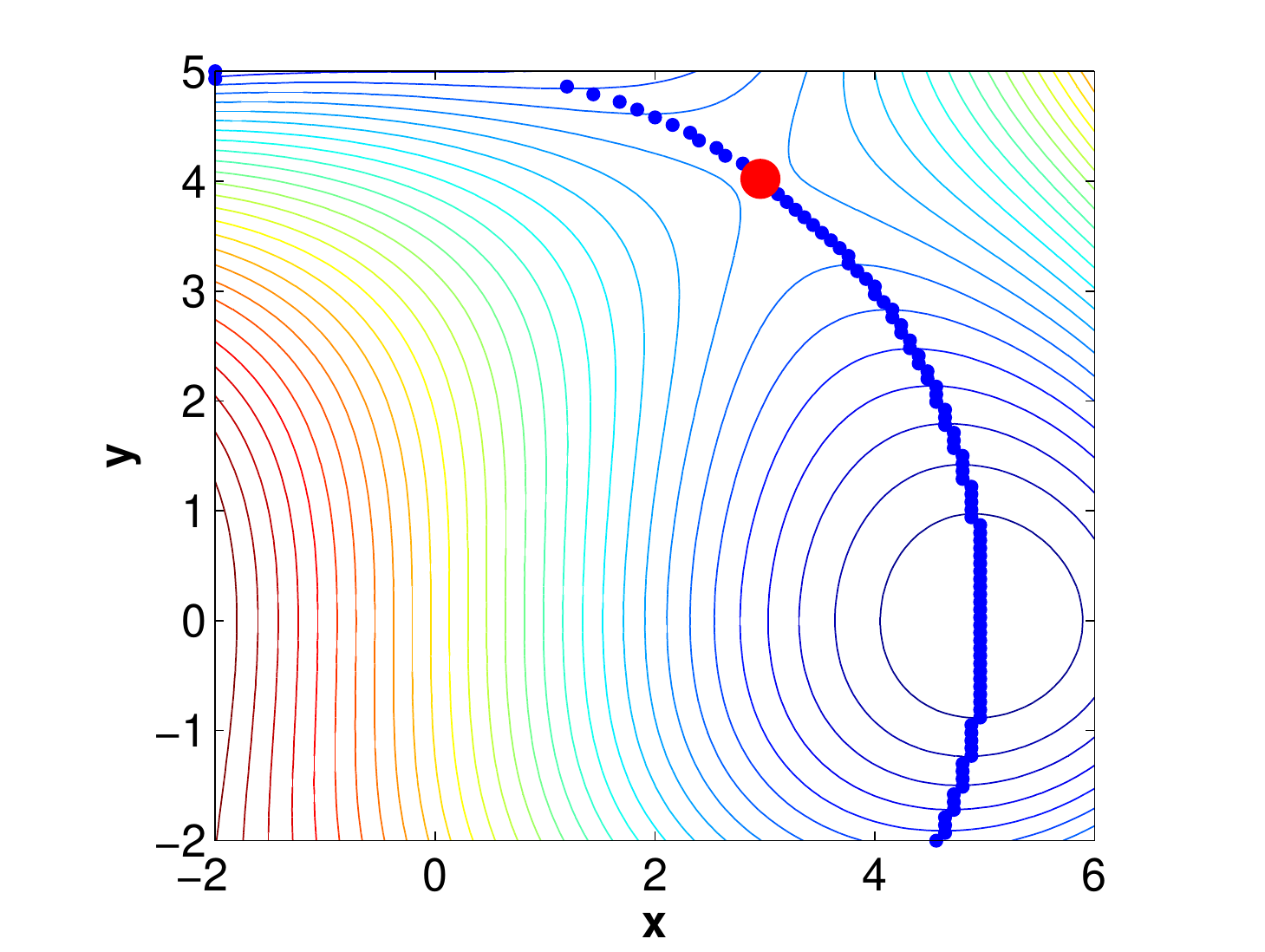}}
\caption{(a,b) Contour plots of the function function $f(x,y) = \sin (x) \cos (y)$ in the range $x,y \in [-2\pi,2\pi]$. The maxima are at the center of the red circles, the minima are at the center of the blue circles, and the saddle points are at the intersections of the green lines. The straight bold lines overlapped with the contour plots are the path taken by an algorithm which attempts to find a saddle point of $f(x,y)$, by starting at point $(-0.1 , \pi-1.5)$ (red asterisk), and (a) successively minimizing in the $x$ direction and maximizing in the $y$ direction, and (b) successively minimizing in the direction which makes an angle of $\pi/10$ with the $x$ axis and maximizing in a direction which makes an angle of $\pi/10$ with the $y$ axis. (c) Contour plot of the function $f(x,y) = 2x^3+6xy^2-3y^3-150x$ in the range $(x,y) \in [-2,6] \times [-2,5]$. The small (blue) dots represent minima of $f(x,y)$ with respect to the $x$ direction and the large (red) dot represents the maximum of these minima, which is a good approximation of a saddle point.}
\label{fig-saddle_testfunction}
\end{figure*}

In order to better understand why this method may fail to find saddle points, one can consider a much simpler example, the function $f(x,y) = \sin (x) \cos (y)$. This function is periodic, oscillates between -1 and 1, and has infinitely many minima, maxima and saddle points (Fig.~\ref{fig-saddle_testfunction}). Let us now assume that we want to find the saddle points of $f(x,y)$ using the method of minimizing and maximizing several times alternating between the two, using $x$ as minimizer and $y$ as maximizer. If we start with a non-saddle point, the algorithm fails to find a saddle point and instead, after 2 ou 3 steps, alternates between absolute minima and maxima. Even if we start at a saddle point, successive iterations will move away from it and back to jumping between maxima and minima (see Fig.~\ref{fig-saddle_test_0}). No initial point will allow this algorithm to converge to a saddle point, no matter how close it is to it. This is precisely what happened with the case of $E_{\text{HF}}$. Depending on the pair $(n,U)$ in question, $E_{\text{HF}}$ may or not behave in a way that allows saddle points to be found using this method.

One possible alternative approach to finding saddle points of $f(x,y)$ is to attempt to extremize the function in a different direction (or rotating the axes, which is another way to see it). Instead of using $x$ and $y$ as variables, we can use two auxiliary variables which are linear combinations of $x$ and $y$ but still orthogonal. For instance, we can alternate between the following two:
\begin{itemize}
\item replace $x \rightarrow x_0 +r \cos \theta$, and $y \rightarrow y_0 +r \sin \theta$, and minimize with respect to $r$;
\item replace $x \rightarrow x_0 +r \sin \theta$, and $y \rightarrow y_0 -r \cos \theta$, and maximize with respect to $r$.
\end{itemize}

Setting $\theta = 0$ would reduce this to simply minimizing with respect to $x$ and maximizing with respect to $y$, as described above. Due to the symmetry of the simple function we used, we should ideally use $\theta = \pi/4$ for the fastest convergence. For angles close enough to $\pi/4$ we can indeed find a good approximation for  a saddle point. Nevertheless, if we deviate too much from $\pi/4$, saddle points may not be found anymore. Using a "wrong" angle will make the algorithm diverge even if we start very close to a saddle point, even if the starting point is a saddle point itself. The algorithm will often be circling the saddle point (see Fig.~\ref{fig-saddle_test_pi20}).

The conclusion is that alternating between maximization and minimization is too sensitive to the initial point to be used to find saddle points. In the case of the simple highly-symmetric function $f(x,y)$, the saddle point is located at the center of the squares or rectangles that are drawn when connecting the points given by the algorithm, but this was shown to not be the case with $E_{\text{HF}}$. Moreover, this ideal angle depends on the pair $(n,U)$ which is being studied, and some angles may not even produce closed polygons, but rather diverge to infinity in a certain direction.

The working alternative is failproof in the sense that it will always converge to one saddle point. It consists of calculating the value of $f(x,y)$ for many points in an area which is known to contain at least one saddle point, making a list of the maximum value of the function for each $x$ and then finding the minimum of all the maxima that were found. If $f(x,y)$ is continuous we will end up on a saddle point. In the case of $E_{\text{HF}}$, we have not two variables but three: $m_A$, $m_B$ and $\delta_A$. In order to achieve an acceptable precision, it would be reasonable to calculate, say,  100 values of $E_{\text{HF}}$ in each direction ($m_A$, $m_B$, and $\delta_A$), for a total of $10^6$ values for each pair $(n,U)$ in our phase diagram, which could take a long time and would have a precision of about two decimal places. A more efficient alternative is to divide each direction into fewer parts, say 7 or 8, finding an initial approximation to the saddle point, and then repeating this a dozen times considering a smaller hypercube centered on the new point. Assuming each direction is divided into 7 parts (i.e. we calculate 8 function values in each direction) and the side of the hypercube is halved with each of 12 iterations, we now calculate a total of $7^3 \times 12 \approx 4000$ values of $E_{\text{HF}}$ per pair $(n,U)$. This was the procedure we used, which has a precision of around $1/4096$, or three decimal places. An illustration of this method is shown in Fig.~\ref{fig-saddle_test_success}, where we have used the function $f(x,y) = 2x^3+6xy^2-3y^3-150x$ and successfully found the saddle point $(3,4)$.

\section*{Acknowledgements}

R. G. Dias acknowledges the financial support from the Portuguese Science and Technology Foundation (FCT) through the program PEst-C/CTM/LA0025/2013. J. D. Gouveia acknowledges the financial support from the Portuguese Science and Technology Foundation (FCT) through the grant SFRH/BD/73057/2010.

\bibliography{helix1}
\bibliographystyle{unsrt}
\end{document}